\documentclass[final,5p,times,twocolumn]{elsarticle}

\usepackage{graphicx}
\usepackage{subfigure}
\usepackage{multirow}
\usepackage{wrapfig}

\usepackage{tabularx}
\usepackage{booktabs}

\usepackage[utf8]{inputenc}
\usepackage{amsmath}
\usepackage{amsthm}
\usepackage{amsfonts}
\usepackage{epsfig}
\usepackage{psfrag}
\usepackage{pstricks}
\usepackage{algorithm}
\graphicspath{{./Figures/}}

\usepackage{amssymb}
\usepackage{bm}

\biboptions{comma,sort&compress}

\usepackage{color}
\usepackage{ulem}
\newcommand{\rev}[2]{#2}

\journal{Computer-Aided Design}

\begin{document}

\begin{frontmatter}

\title{Variational Direct Modeling: A framework towards integration of parametric modeling and direct modeling in CAD}

\author[zju]{Qiang Zou\corref{cor}}\ead{qiangzou@cad.zju.edu.cn}
\author[ubc]{Hsi-Yung Feng}
\author[zju]{Shuming Gao}

\cortext[cor]{Corresponding author.}
\address[zju]{State Key Laboratory of CAD$\&$CG, Zhejiang University, Hangzhou, 310027, China}
\address[ubc]{Department of Mechanical Engineering, The University of British Columbia, Vancouver, BC, V6T 1Z4, Canada}

\begin{abstract}
Feature-based parametric modeling is the de facto standard in CAD. Boundary representation-based direct modeling is another CAD paradigm developed recently. They have complementary advantages and limitations, thereby offering huge potential for improvement towards an integrated CAD modeling scheme. Most existing integration methods are developed by industry and typically treat direct edits as pseudo-features, where little can be said about seamless integration. This paper presents an alternative method for seamless parametric/direct integration, which allows parametric and direct edits to work in a unified way. The fundamental issues and challenges of parametric/direct integration are first explained. A framework is then proposed to handle those information inconsistencies, based on a detection-then-resolution strategy. Algorithms that can systematically detect and resolve all possible types of information inconsistencies are also given to implement the framework. With them, model validity can be maintained during the whole model editing process, and then the discrepancy between direct edits and parametric edits can be resolved. The effectiveness of the proposed approach has been shown with a series of case studies and comparisons, based on a preliminary prototype.
\end{abstract}

\begin{keyword}
Computer-aided design \sep Solid modeling \sep Parametric/Direct integration \sep Information inconsistency \sep Validity maintenance \sep Decision-making \sep Intelligent CAD

\end{keyword}

\end{frontmatter}

\section{Introduction}
\label{sec:introduction}
Computer-Aided Design (CAD) has been the dominant industrial practice for product design. One of its primary usages is to create and edit computer models of products \cite{li2020survey,sapidis2007geometric}. CAD modeling so defined has experienced three broad stages of development: solid modeling, parametric modeling, and direct modeling. Solid modeling began in the 1970s with two competing approaches \cite{hoffmann2005constraint}: constructive solid geometry (CSG) and boundary representation (B-rep). They have respectively motivated parametric modeling in the late 1980s and direct modeling in the late 2000s.

The CSG approach represents a solid as successive combinations (via regularized Boolean operations) of primitive shapes, \cite{requicha1985boolean,requicha1982solid,requicha1983solid}. Parametric modeling inherits this procedural way of working but introduces features to replace the primitive shapes \cite{camba2016parametric,gonzalez2017survey}. By doing so, it allows easy embedding of associativity\footnote{Associativity means that geometric entities of a CAD model are associated through, for example, geometric constraints \cite{shah1998designing}.} into shapes, which in turn yields benefits of automatic change propagation, shape reuse, and design intent embedding \cite{shah1998designing}.  Nevertheless, associativity also introduces complexity and rigidity into CAD models. Most notably, a parametric model's construction history must be carefully thought out before modeling because this will restrict the user to what can be edited once the model is built. Making changes to an existing parametric model requires a good understanding of its construction history and the mathematics behind. For these reasons, users often find parametric modeling hard to learn and use \cite{ElHani2012,monedero2000parametric}.

The B-rep approach represents a solid by specifying the boundary between the solid and void. The boundary is stored as a collection of faces sewed together \cite{requicha1982solid}. Direct modeling allows users to directly manipulate those boundary faces to attain solid variants \cite{Zou,qin2021automatic}. This way of working provides benefits of flexible edits, fast update, and intuitive interaction. It is a flexible approach because, in principle, it can change a model to any shape. Model updates are fast since only the modified boundary faces need to be updated and the majority remain unchanged. The intuitive interaction implies a shallow learning curve. However, these advantages come at a high price: associativity information is lost, and parameter-driven automatic change propagation is no longer available \cite{Fu2017}.

Clearly, parametric and direct modeling have complementary advantages and disadvantages. Their integration can thus provide both strengths, i.e., automatic change propagation and high modeling flexibility. Nevertheless, integrating them is not trivial. The fundamental issue lies in the information inconsistencies caused by parametric/direct edits when they are applied to a same CAD model (a detailed explanation on this will be given in Section~\ref{sec:from-integration-to-inconsistency}). Without being consistent with each other, information in a CAD model will produce an invalid model, e.g., a non-solid and/or an over-constrained model.

To date, there have been five integration strategies reported by industry and academia (as will be detailed in Section \ref{sec:relatedwork}). When dealing with the above fundamental issue of information inconsistencies, they consistently try to convert direct edits into certain feature operations, where lossy conversions can happen and little can be said about seamless integration. By seamless we mean the integration can provide the same editing capability and modeling behavior as in unintegrated modelers. In other words, if direct edits are applied, the model acts like a B-rep model; if parametric edits are considered, it acts like a parametric model.

This paper presents our attempts on seamless parametric/direct integration. The fundamental problem to be studied is resolving the information inconsistencies in a CAD model caused by parametric/direct edits. The primary goal lies in systematically detecting and resolving all possible types of information inconsistencies so that the following three technical requirements can be met during all modeling operations: model validity (being solid and well-constrained); continuous model shape variations; and minimal model constraint changes. Section \ref{sec:from-inconsistency-to-decision-making} will provide detailed discussions on the formulation of these three requirements.

The major contributions of the paper include:
\begin{enumerate}
    \item Identifying the research problem underlying parametric/direct integration and its fundamental challenges;
    \item Proposing a framework, called variational direct modeling, that can systematically handle any information inconsistencies in a CAD model after parametric/direct edits; and
    \item Formulating the high-level tasks of detecting and resolving information inconsistencies as well-defined technical problems such as Booleans on solids and over-constraint maximization.
\end{enumerate}

The following sections begin with a review on parametric modeling, direct modeling, and their integration in Section \ref{sec:relatedwork}. A thorough analysis of issues and challenges of parametric/direct integration is given in Section \ref{sec:challenges}. The overall framework of the proposed integration method is presented in Section \ref{sec:Methodology-overview} and elaborated in Section \ref{sec:Methodology-details}. Application examples and comparisons with leading commercial CAD modelers are provided in Section \ref{sec:results}, followed by conclusions on the method's advantages and limitations in Section \ref{sec:conclusion}.

\section{Related work}
\label{sec:relatedwork}

\begin{figure*}[!ht]
    \centering
    \includegraphics[width=0.9\textwidth]{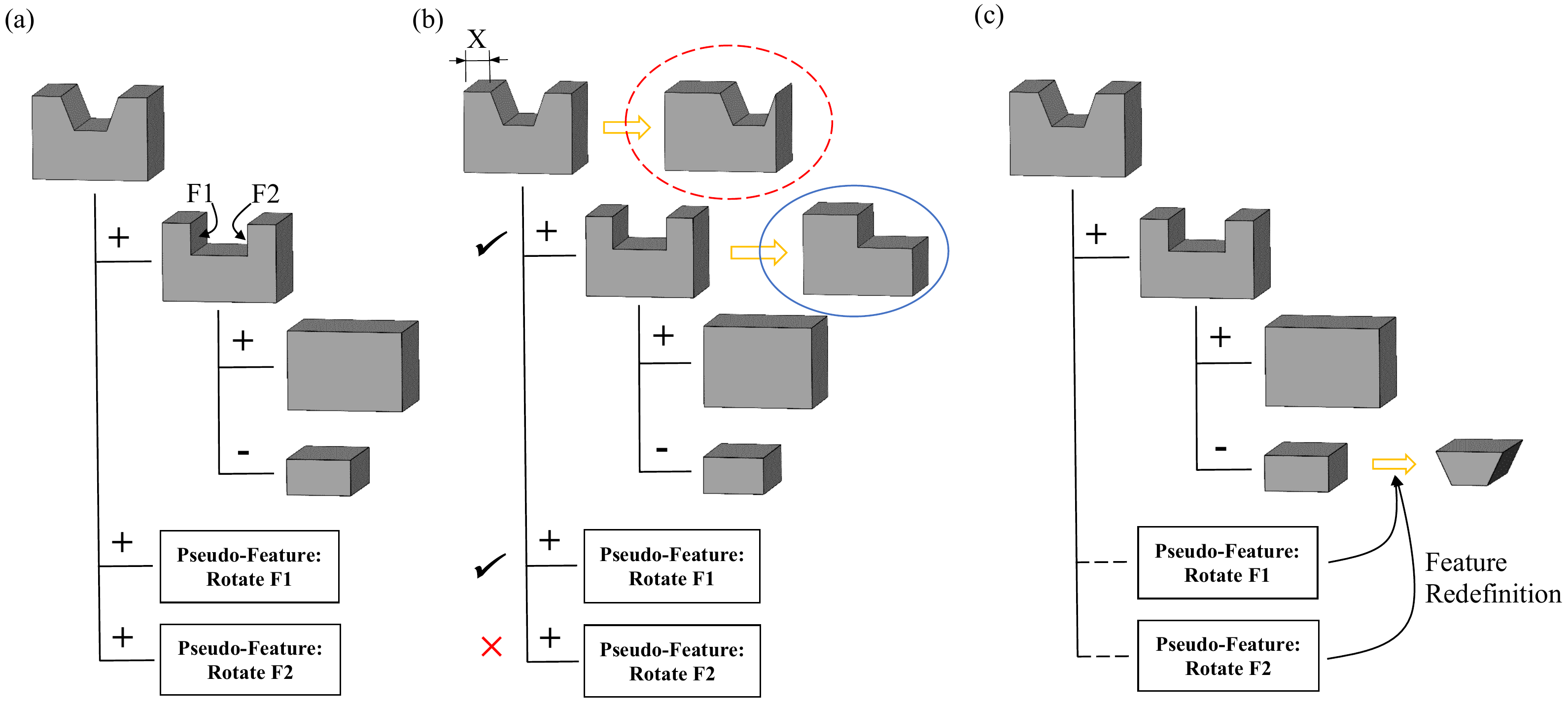}
    \caption{Limitations of the pseudo-feature approach: (a) the model's construction history; and (b) increasing the X dimension cannot give the desired model depicted in the dashed red circle but leads to a history regeneration failure because of the loss of face F2 (as indicated by the model in the blue circle); and (c) a reasonable integration strategy is to redefine the slot feature based on the direct edits to avoid the situation of (b).}
    \label{fig:parametric-messup}
\end{figure*}

Both parametric and direct modeling are built on top of solid modeling, a CAD modeling approach established by Requicha and Voelcker, then at University of Rochester, and Braid and Lang, then in Cambridge \cite{braid1975synthesis,voelcker1977geometric}. Solid modeling was proposed as an improvement to the preceding wireframe and surface modeling paradigms such that the informational incompleteness problem observed in the two paradigms can be solved \cite{Shapiro2002}. It is very suitable for documenting completed designs but lacks the flexibility needed for design modifications.

In view of the above limitation, parametric modeling augments solid models with associativity (in the form of features and constraints) such that changes can be propagated automatically \cite{shah1998designing,kyratzi2022integrated}. With automatic change propagation come design intent embedding and geometry reuse. This shifted CAD from an instance modeler to an ``electronic master" modeler. Nevertheless, associativity also implies complexity and rigidity. Working with a parametric model requires a good understanding of its modeling history \cite{camba2016parametric}. Editing a parametric model is hard, even impossible without rebuilding the model, if the desired changes fall outside the designed parametric family \cite{Raghothama2002}. Updating a parametric model is often time-consuming due to the always ``start-over'' model regeneration mechanism. Maintaining the semantics of a model's feature during all modeling operations is not straightforward \cite{bidarra2000semantic}. All of these make parametric modelers hard to learn and use \cite{ElHani2012,monedero2000parametric}.

Direct modeling emerged recently as a solution to the above limitations of parametric modeling, particularly the rigidity issue. It allows users to directly manipulate the model geometry without considering how the model was built, using simple operations like grab, push, and pull \cite{Tornincasa2010,Ault2016}. Although initially proposed by industry, direct modeling may be traced back to the local operations developed by academia in the 1980s \cite{Grayer1980,Rossignac1990,Stroud2006}. Some widely used local operations in today's CAD systems are filleting and offsetting. The very local operation that direct modeling stemmed from is the tweaking operation \cite{Grayer1980}. It is the predecessor of the push-pull operation in direct modeling. The essential advance made is: tweaking does not allow any violations to the model's topology, while push-pulls allow such violations. With this relaxation, direct modeling achieves unprecedented modeling flexibility. This, however, comes at a high price: associativity is discarded, and direct modeling lacks the capability of parameter-driven modifications. 

Some efforts have been made to allow users to have both direct and parametric capabilities in the same modeler. To the best of the authors' knowledge, there are five reported parametric/direct integration approaches, as summarized below. 

\textbf{Integration Method I: Pseudo-Features} \quad This approach, proposed by industry, simulates direct modeling within a history-based parametric modeler. User-specified direct edits are simply added to the end of the model's construction history as pseudo-features, and the original history remains exactly as before \cite{bettig2011geometric,Fu2017}. Fig.~\ref{fig:parametric-messup}a illustrates this strategy (this figure was created based on the modeling behavior of Siemens NX). This approach is very easy to implement and thus has been adopted by most CAD vendors. However, simply adding direct edits to the end of the model's history creates complex modeling history with questionable parameters \cite{bettig2011geometric}. This could mess up the model's parametric information and lead to loss of meaningful parametric controls. The example shown in Fig.~\ref{fig:parametric-messup}b illustrates this situation: changing dimension $X$ causes an unpredictable loss of face F2, which in turn disrupts the model regeneration process at the second pseudo-feature. The perfect solution to this problem is not using pseudo-features but transforming user-specified direct edits into appropriate redefinition of relevant features, as illustrated in Fig.~\ref{fig:parametric-messup}c. The present work serves to address the issue using this strategy.

\textbf{Integration Method II: Mode Switching} \quad This approach, proposed by industry as well, allows users to switch between direct and parametric modeling modes in a same CAD software package. Its implementation is very simple. When switching from the parametric mode to the direct mode, the parametric model is downgraded to a B-rep model, then direct modeling becomes applicable. However, it cannot recover any parametrics when switching back from the direct mode to the parametric mode \cite{nag2015methods}.

\textbf{Integration Method III: Synchronous Technology} \quad This approach is, again, proposed by industry or more specifically by Siemens NX. It can be viewed as as an improvement to the above mode switching method. Instead of convering the whole feature model to a B-rep model when switching from the parametric mode to the direct mode, it does a partial conversion. The basic algorithm behind the partial conversion is as follows. Features are separated into direct-edit features and ordinary features. Only direct-edit features are converted to a B-rep model. When modeling, the user moves an ordinary feature to the direct-edit feature set and then carries out direct modeling, but this method requires all ordinary features prior to the feature being moved in the modeling history to be moved as well, regardless of any design intent loss \cite{Chad2008}. Clearly, this causes unnecessary loss of parametrics. (Note that there was a time when synchronous technology also referred to 3D variational modeling \cite{lin1981variational,chung2000framework}, but this interpretation appears not have been implemented in the current Siemens NX software package.)

\textbf{Integration Method IV: Operation Translating} \quad This approach, proposed by both academia and industry (Autodesk), translates direct edits into operations of parameter tuning and/or order rearrangement of the features already presented in the model's construction history \cite{Fu2017,qin2021automatic}. Currently, the translation is done by using heuristics, e.g., \cite{qin2021automatic}. This way of working may help but cannot solve the problem altogether because not all direct edits are achievable through those feature operations. To make matters worse, feature parameter tuning for a given direct edit (if achievable) is often not unique. How to generate an exhaustive list of parameter tuning options and make robust decisions among them still remains unknown.

\textbf{Integration Method V: Constrained Direct Modeling} \quad This approach, again proposed by industry, allows users to apply direct edits while keeping all geometric constraints of the model in the background \cite{ushakov2008variational}, through efficient geometric constraint solving \cite{hoffman2001decomposition}. The geometric constraints being kept are generated by automatic constraint recognition algorithms, e.g., \cite{cordier2013inferring}. The disadvantage of this method is that the recognized geometric constraints usually differ from the original design intent. In addition, this method downgrades direct modeling to merely a graphical user interface (GUI) tool for making parametric modifications. For this reason, although the method also takes the name of variational direct modeling, it is essentially another version of the traditional variational modeling and has little to do with direct modeling, thereby differing substantially from the present work.

The above review suggests that substantial progress has been made in understanding and implementing parametric and direct modeling, and several directions have been tried with the goal of integrating them. The reported integration methods are very inspiring to this work. Nevertheless, their developments are still at the early stage, and there are inherent drawbacks if seamless parametric/direct integration is desired. This work follows this research direction but uses a new way to approach  seamless parametric/direct integration. More specifically, an additional module, called variational direct modeling, is to be proposed to maintain the information consistency in a model undergoing parametric/direct edits. Different from current conversion-based methods which work in an indirect manner, it will resolve any possible information inconsistencies in a straight and systematic manner.

\section{Issues and challenges of parametric/direct integration}
\label{sec:challenges} 
As already noted, for parametric/direct integration to be seamless, the model should act like a B-rep model for direct edits, and a parametric model for parametric edits. To achieve this goal, the underlying problem needs to be solved is: parametric/direct edits cause information inconsistencies among topology, geometry, and constraints in the model. The fundamental challenge of resolving such information inconsistencies is: there often exist many resolution options, and decision-making among them needs to be systematic. The next two subsections explain these two statements.

\subsection{From integration to information inconsistency}
\label{sec:from-integration-to-inconsistency} 
There are three basic types of information in a CAD model: geometry, topology, and geometric constraint system (GCS). Geometry and topology together determine the actual shape of the CAD model; the GCS wrapped around them represents the associativity embedded into the CAD model. Parametric edits are made through the GCS information layer, and direct edits via the geometry information layer. The major issue here is: when an information layer is edited, the changes cannot be automatically reflected in others by current model representation schemes. As a result, the consistency of the three information layers in the pre-edit model is broken, and then an invalid model is generated after a parametric/direct edit.

\begin{figure*}[hbt!]
    \centering
    \includegraphics[width=0.7\textwidth]{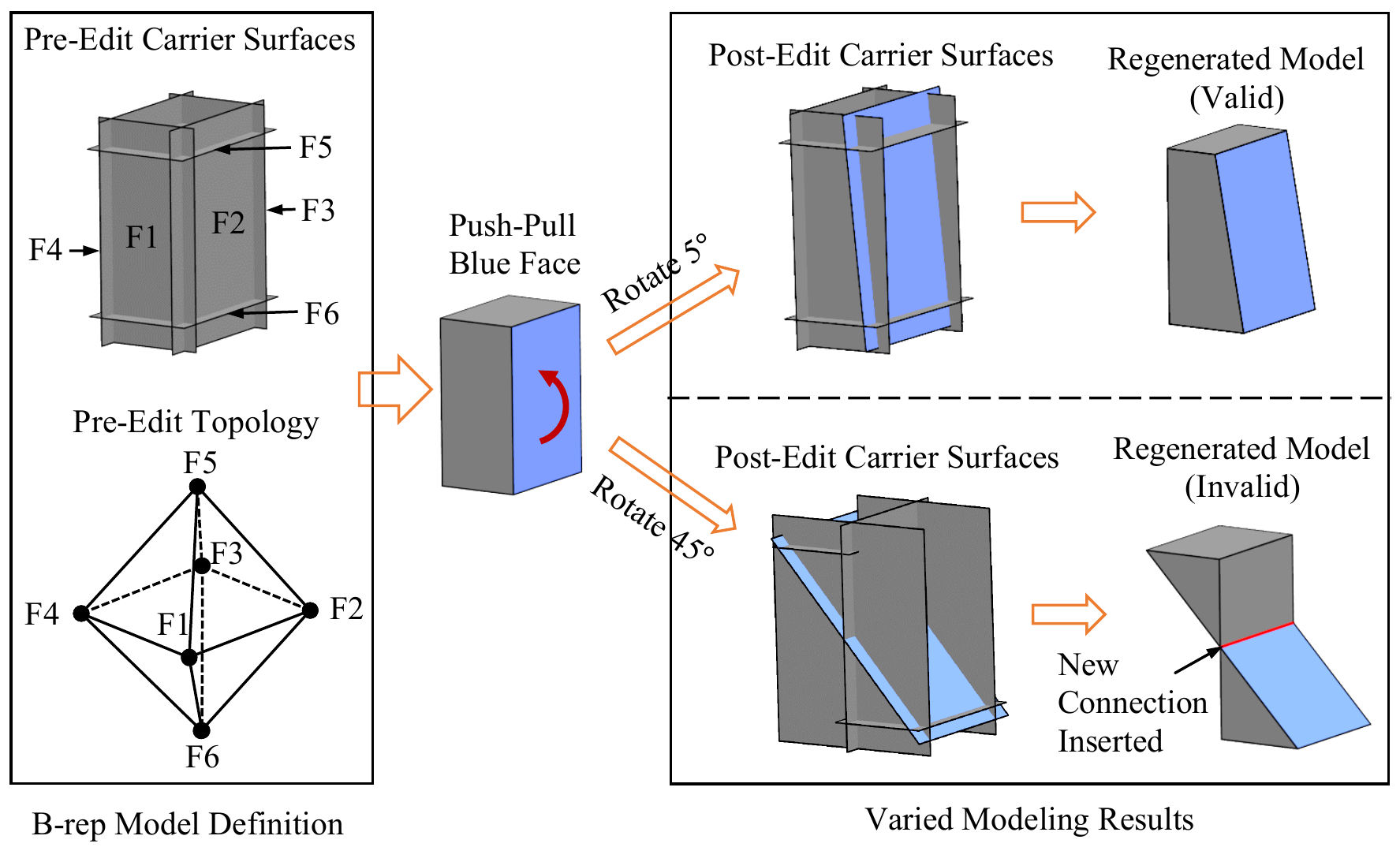}
    \caption{Examples of geometry-topology inconsistencies (blue face: push-pulled face; red arrow: push-pull direction). Model regeneration means boundary re-evaluation using the post-edit surfaces and pre-edit topology.}
    \label{fig:boundary-regeneration}
\end{figure*}

\begin{figure*}[hbt!]
    \centering
    \includegraphics[width=0.75\textwidth]{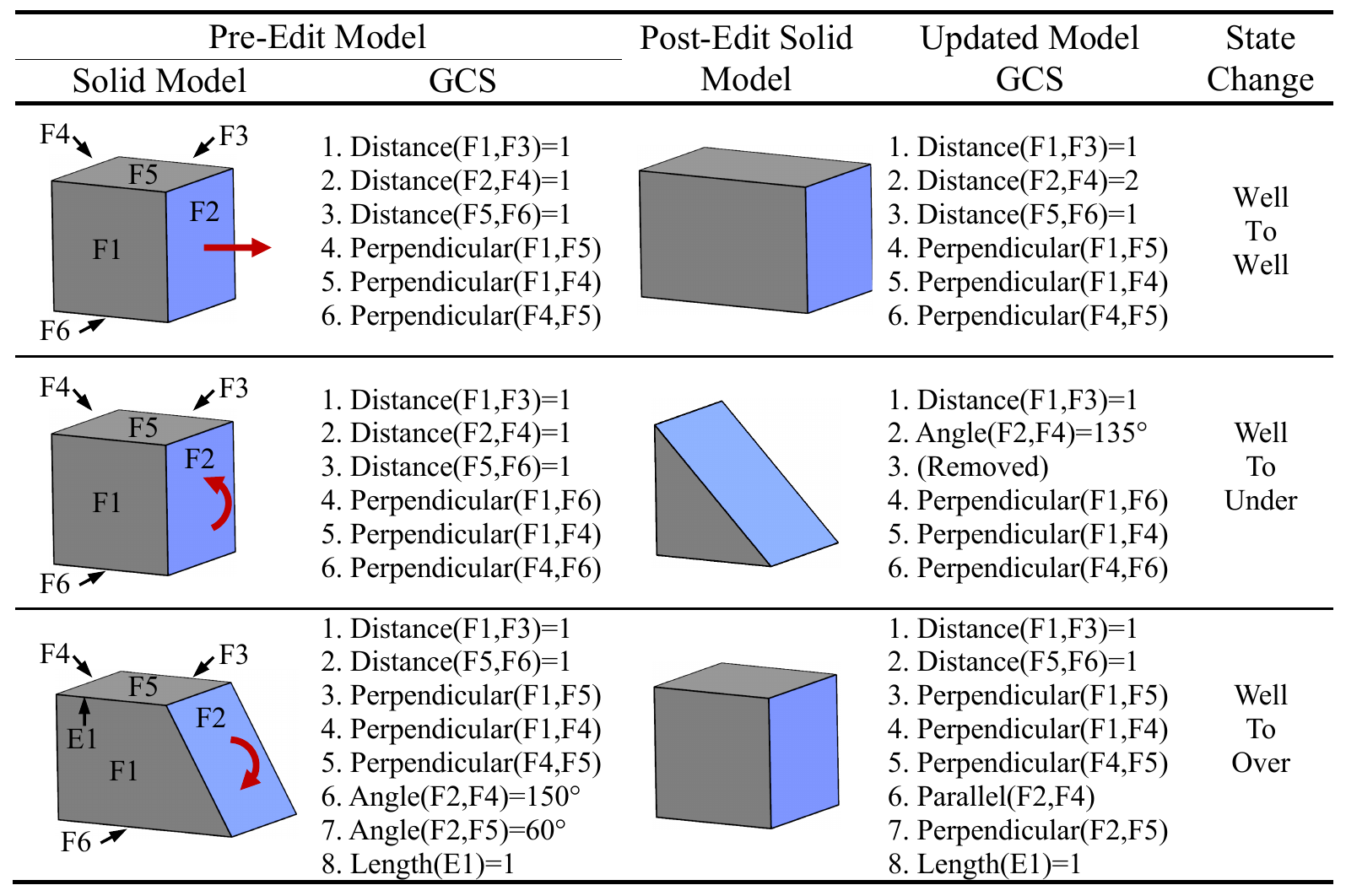}
    \caption{Model GCS update and varying constraint state change results (blue face: push-pulled face; red arrow: push-pull direction) \cite{zou2020decision}. Note that the first two columns underneath ``Pre-Edit Model" are considered a single column.}
    \label{fig:constraint-update}
\end{figure*}

Three types of information inconsistency are possible:
\begin{enumerate}
    \item \textbf{Geometry-topology inconsistency (GTI)} \quad After a direct edit, the changed geometry could be inconsistent with the unchanged topology in two ways: (1) some connections in the topology cannot be formed any more, e.g., two originally connected planar faces become parallel after a push-pull; and (2) extra connections are formed, as shown in Fig.~\ref{fig:boundary-regeneration}. Either way leads to invalid boundary faces \cite{Zou}, e.g., open face, self-intersected face, and non-manifoldness. Such faces break the validity of the pre-edit model. (For formal definitions of the validity of B-rep solid models, please refer to Mantyla's work \cite{Mantyla1984a} or our previous work \cite{Zou}.) 
    \item \textbf{Shape-associativity inconsistency (SAI) A} \quad After a direct edit (and successfully resolving GTIs), the model takes a new shape. This shape, or a portion thereof, will have new boundary faces and dimensions, making some geometric constraints in the model GCS inapplicable any more. Take the model in the second row of Fig.~\ref{fig:constraint-update} as an example. Before the rotational push-pull move, it is a cuboid with six geometric constraints. After the move, its top face is gone, and the right face is slanted. \rev{Then all geometric constraints related to these two changed faces become disagree with the new shape, necessitating constraint update.}{Then all geometric constraints related to these two changed faces become incompatible with the new shape, requiring a constraint update.} Updating model GCS with the new model shape will remove some existing constraints from and/or add new constraints to the model GCS, which could break the model's well-constraint state. Take the same model as an example. After constraint update according to the new shape (i.e., a prism), the second geometric constraint takes a new parameter value, and the third one should be removed. Consequently, an under-constrained GCS is generated. The first and third rows of the figure show two GCS update cases where the constraint state remain unchanged and the constraint state changes from well-constrained to over-constrained, respectively.
    \item \textbf{Shape-associativity inconsistency (SAI) B} \quad After a parametric edit, some parameters of the model GCS are changed, and then the consistency between the original model shape and GCS is broken.
\end{enumerate}

The third inconsistency type can be handled using existing research work, whereas the first two require new input. For the third type, the essential task involved is to solve the model GCS with the given new parameter values, and then to update the model shape accordingly. Geometric constraint solving is a well-established field, and many effective algorithms have been made available \cite{bettig2011geometric,hu2017over}. \rev{For this reason, this work will not focus on the third type. SAI will then refer exclusively to SAI (A) in the following text, unless otherwise stated. It should be noted here that, as revealed by a recent study by Gonzalez-Lluch et al. \cite{gonzalez2021constraint}, a user-specified GCS often contains redundant constraints which can hinder design reusability. It is thus necessary to eliminate such redundancies before handling SAI (B), using tools like the one presented in \cite{gonzalez2021constraint}.}{It should be noted here that, as revealed by a recent study by Gonzalez-Lluch et al. \cite{gonzalez2021constraint}, a user-specified GCS often contains redundant constraints which can hinder design reusability. It is thus necessary to eliminate such redundancies before handling SAI (B), using tools like the one presented in \cite{gonzalez2021constraint}. For this reason, this work will not focus on the third type. SAI will then refer exclusively to SAI (A) in the following text, unless otherwise stated.}

\rev{constraint redundancy in profiles significantly and unnecessarily compromises model conciseness, robustness, and the overall model quality, which negatively affects user productivity and downstream processes.}{}

To summarize, the changed geometry due to a direct edit could cause inconsistencies with the unchanged topology and constraints, consequently generating an invalid solid model and/or GCS. For GTIs, the cause is either losing old connections or inserting new connections. They take the form of invalid boundary faces. For SAIs, the source is due to situations where there are extra constraints or insufficient constraints to restrict the degree-of-freedoms (DOFs) of the model geometry. They take the form of over-constraint \rev{}{(including redundant constraints)} or under-constraint.

\subsection{From information inconsistency to decision-making}
\label{sec:from-inconsistency-to-decision-making}
\begin{figure}[t]
    \centering
    \includegraphics[width=0.45\textwidth]{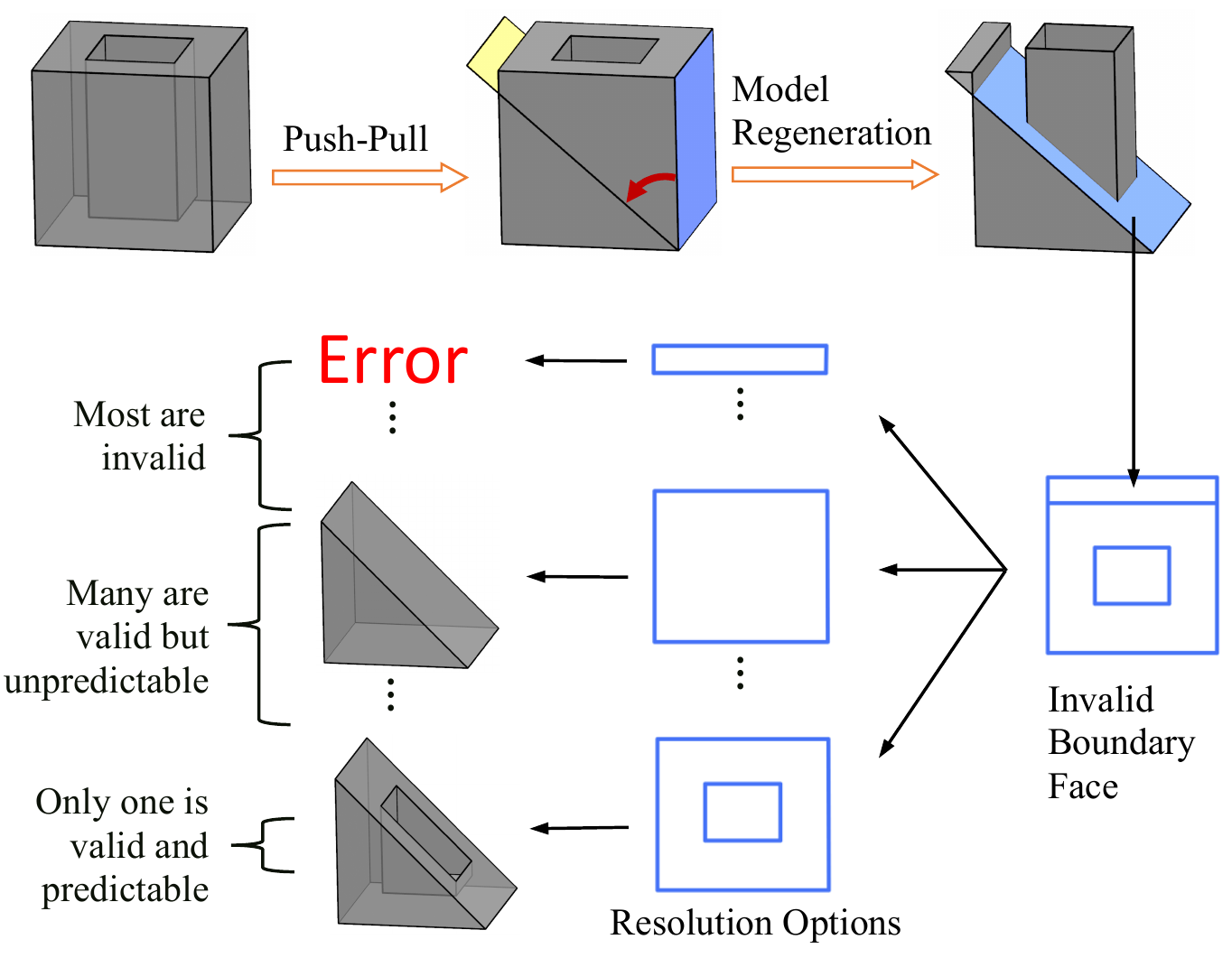}
    \caption{Examples of different GTI resolution options and their varying outcomes.}
    \label{fig:resolution-options-gti}    
\end{figure}

GTIs and SAIs require rectifications on the model topology and GCS to accommodate the changed geometry. Resolving information inconsistencies is not trivial because there often exist many resolution options. Fig.~\ref{fig:resolution-options-gti} shows one such example for GTIs. In this example, a rotational push-pull direct edit is applied to the blind hole model, where the face colored blue is the push-pulled face, and the orange face indicates the target position. If the original topology is not corrected to accommodate the face's new position, an invalid model will be generated, see the rightmost model at the first row, where self-intersections (indicating model invalidity) are generated. Even for a single invalid boundary face, there are multiple options that can do invalidity resolution and give valid boundary loops, as shown by the three sample options in the second row of the figure. However, decision-making among resolution options for neighboring faces is coupled because, for the final model to be valid, any edge on a resolved face should be shared with a neighboring face so as to satisfy the manifoldness condition, and meanwhile the neighboring faces have to be well-bounded as well. For example, if the top one of the three sample options is chosen, there is no way that other \rev{resultion}{resolution} options on neighboring faces can match with it to give a valid model. As a result, among the many valid resolution options for individual invalid boundary faces, only a few combinations of them can give valid outcomes for the whole model. Most of the combinations will lead to resolution failures. Hence, decision-making in GTI resolution is error-prone. 

In addition to the validity issue, the generation of predictable modeling results is also challenging. Among the many resolution options, some will lead to invalid modeling results; among the valid results, there will only be one in line with the designer's intent. For example, the middle resolution of the three sample options in Fig.~\ref{fig:resolution-options-gti} will lead to a valid overall model but with a missing hole, while the bottom one can give a satisfactory result. It is difficult to always attain intended results due to the involvement of design intent \cite{camba2015assessing}. To make the problem more tractable, this work relaxes the predictability requirement a little bit to the continuity requirement, which means that the model variation follows a continuous change pattern. This relaxation is inspired by Raghothama and Shapiro's work \cite{raghothama1998boundary}. They showed that much, if not all, of the unpredictable modeling behavior in CAD can be avoided by the continuity requirement. (Note that relaxation means approaching a difficult problem by a nearby problem that is easier to solve. In our case, the predictability is the difficult problem, and the continuity is the nearby problem.) Overall, the fundamental challenge of decision-making in GTI resolution lies in the robustness towards generating valid modeling results and continuous model variations.

SAI resolution is under the same situation of multiple resolution options. Consider, for example, the bottom case in Fig.~\ref{fig:constraint-update}. The updated model GCS has an over-constrained part: a cyclical dependency among constraints 5, 6, and 7. Removing any constraint not involved in the dependency cannot resolve the over-constraint and leads to a failed resolution. Only removing a constraint relevant to the dependency can resolve the over-constraint, as shown in Table~\ref{tab:candidate-resolution}. Therefore, robustness towards generating relevant constraints (i.e., valid resolution options) is one of the fundamental challenges of SAI resolution.

\begin{table*}[htb!]
    \caption{Candidate resolution options for the bottom case in Fig.~\ref{fig:constraint-update} \cite{zou2020decision}.} 
    \centering
    \includegraphics[width=0.8\textwidth]{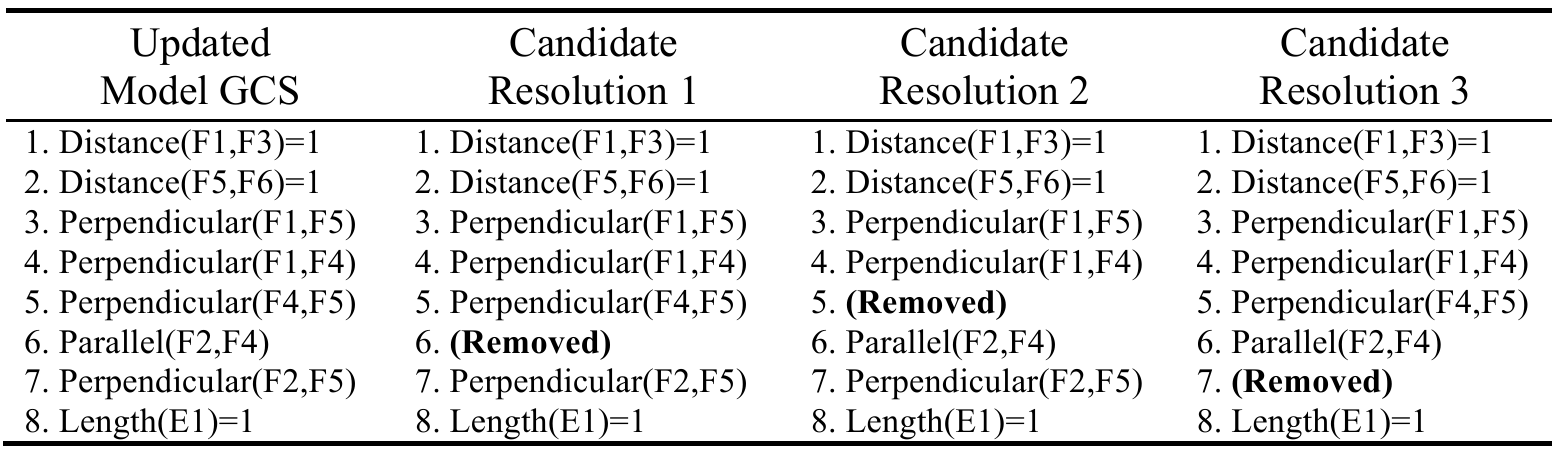}
    \label{tab:candidate-resolution}
    \vspace{-4mm}
\end{table*}

Quite often, there is more than one valid resolution option even if the resolution method successfully avoids invalid resolution options. The options presented in Table~\ref{tab:candidate-resolution} are typical examples for such a situation. Decision-making among these valid resolution options is difficult due to, again, the involvement of design intent. As a result, a completely automatic decision-making method seems to be impossible, at least at present. A decision-support scheme may be a more practical choice. The key here is to have a prioritization of valid resolution options so as to recommend them to users in an incremental manner. An effective prioritization scheme should give a good measure of the impact of removing/adding a constraint on the model shape. The problem is, however, not straightforward since the qualitative operation of removing/adding constraints has no direct connections to model shape changes which are quantitative in nature.

In summary, the major issues of parametric/direct integration are the possible GTIs and SAIs in a model undergoing direct edits. To handle them, there often exist many resolution options, and systematic decision-making is needed. For GTI resolution, the challenges lie in the robustness towards generating valid modeling results and continuous model variations. For SAI resolution, the challenges lie in the robustness towards generating valid resolution options and in the effectiveness of prioritizing them.

\section{The proposed methodology: variational direct modeling}
\label{sec:Methodology-overview}
\begin{figure*}[hbt!]
    \centering
    \includegraphics[width=0.8\textwidth]{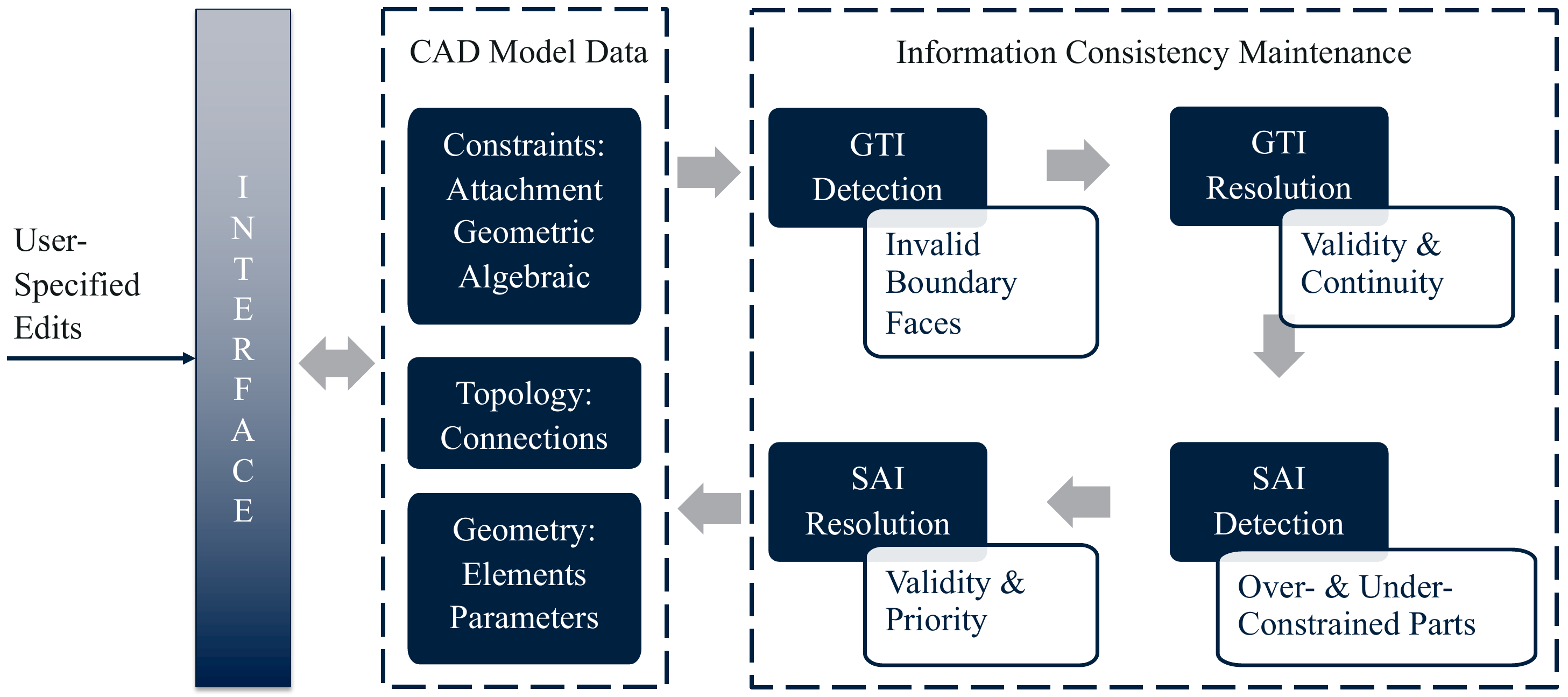}
    \caption{Overall framework of information inconsistency resolution in variational direct modeling.}\label{fig:overall-framework-flow}
\end{figure*}

Based on the above problem analysis, this work proposes a parametric/direct integration framework as shown in Figs.~\ref{fig:overall-framework-flow}, \ref{fig:gti-resolution-framework} and \ref{fig:sai-resolution-framework}. This section focuses on outlining the framework's high-level workflow. The low-level implementation details will be described in the next section. Fig.~\ref{fig:overall-framework-flow} shows the workflow at the highest level. The four modules in the right dashed rectangle represent the framework's main component. It accepts the changed model information (a consequence of the user's edit) as inputs, then corrects the topology (for GTIs) and the constraints (for SAIs) to accommodate the changed model information, and finally outputs a consistent set of information to update the model undergoing edits. The specific model information our method needs to draw from the model is listed in the middle dashed rectangle. The correction basically goes through two steps: detection and resolution. The detection step checks if there are any information inconsistencies and, if yes, prepares the inconsistent information in a form useful for the subsequent resolution step.

\subsection{GTI detection and resolution}
\label{sec:Methodology-overview-gti}
\begin{figure}[htbp]
    \centering
    \includegraphics[width=0.48\textwidth]{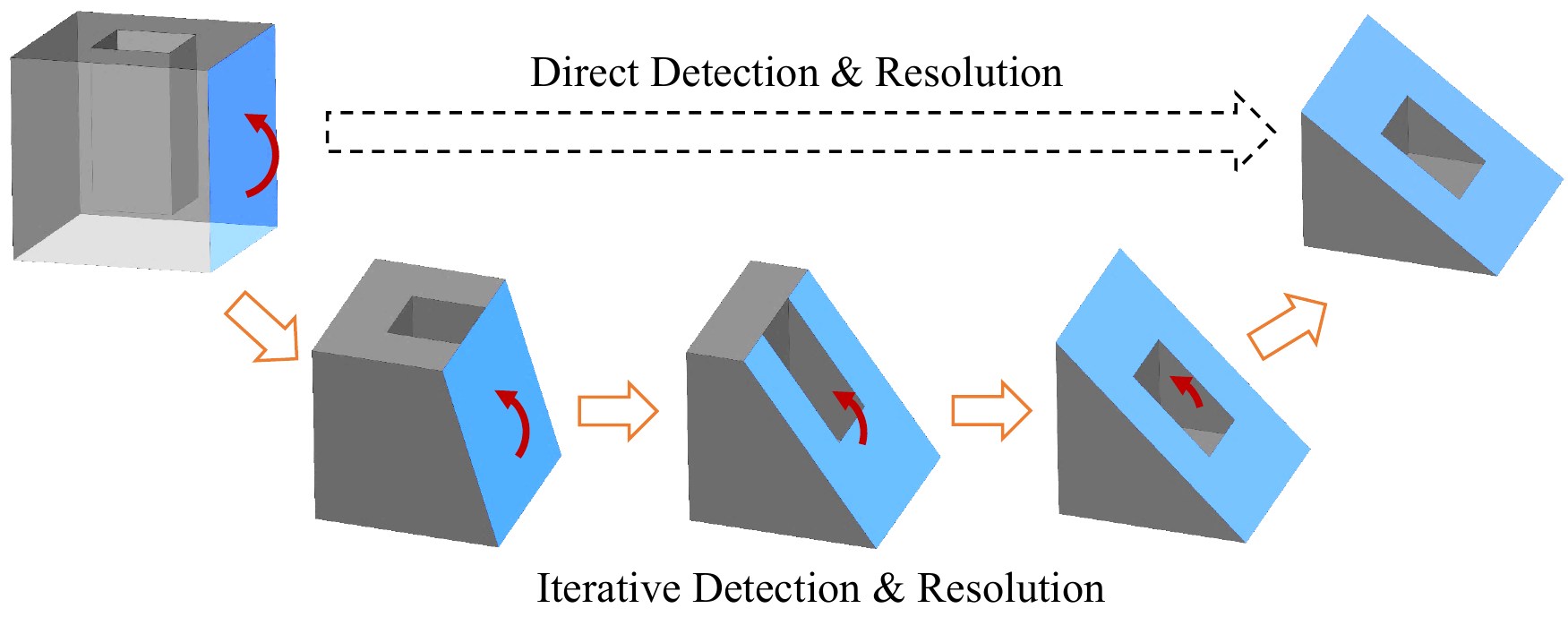}
    \caption{Illustration of iterative GTI detection and resolution (blue face: push-pulled face; red arrow direction: rotation direction; its length: rotation angle).}\label{fig:progressive-gti-framework}
\end{figure}

\begin{figure}[htbp]
    \centering
    \includegraphics[width=0.48\textwidth]{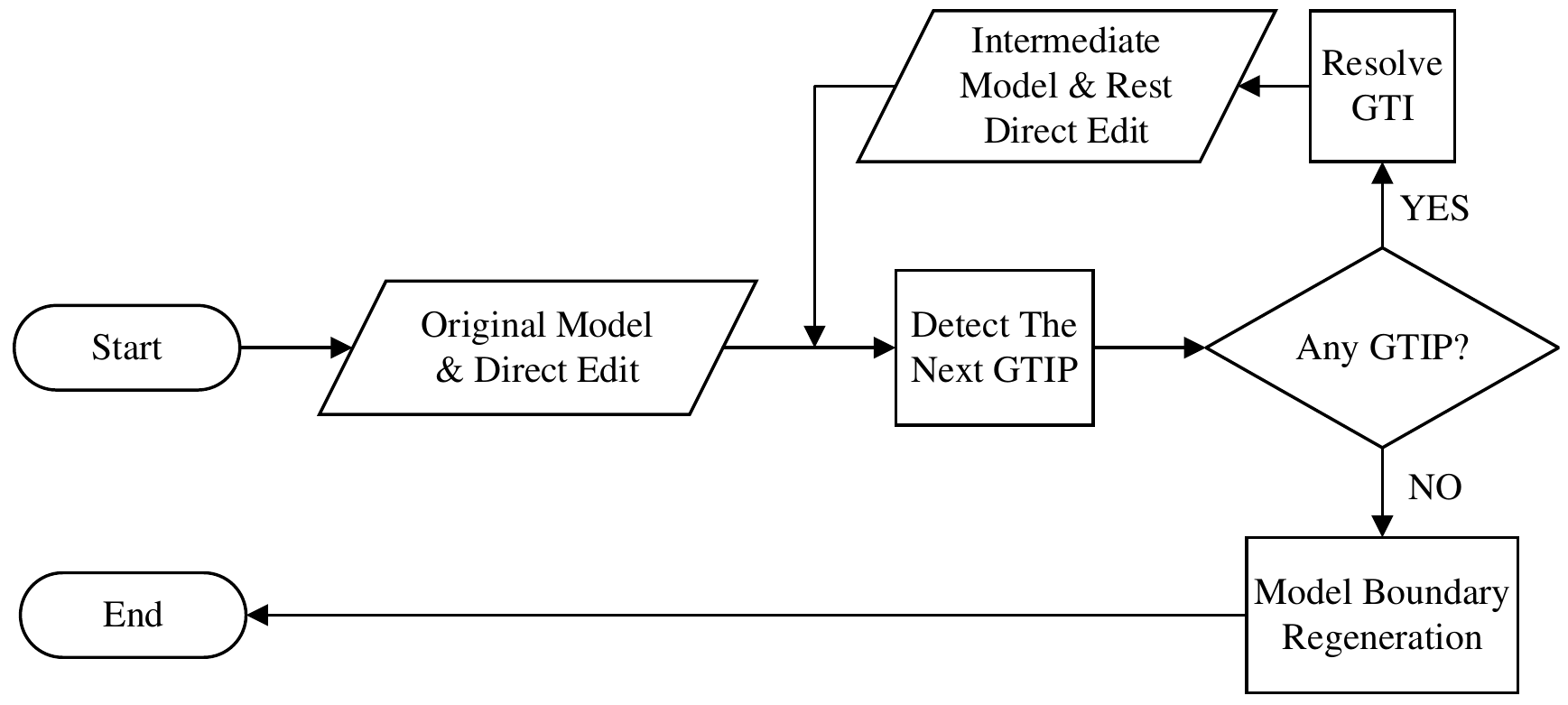}
    \caption{Schematic diagram of GTI detection and resolution.}\label{fig:gti-resolution-framework}
\end{figure}

Instead of postponing GTI detection and resolution to the end of a direct edit, this work employs an iterative approach, repeating the following two procedures until the direct edit is finished: (1) detecting the next point where GTIs occur; and (2) immediately resolving the detected inconsistencies at this point. This approach is illustrated in Fig.~\ref{fig:progressive-gti-framework}, using the model in Fig.~\ref{fig:resolution-options-gti} as an example. There are three critical points where GTIs occur during the entire push-pull move: the first is when the blue face hits the right side face of the hole; the second is when the blue face hits the left side face of the hole; and the last one is when the blue face hits the left side face of the block. Whenever a critical point is reached, the formed GTIs is resolved immediately, generating an intermediate model at this point (see the three models in the lower row of Fig.~\ref{fig:progressive-gti-framework}). The rest of the direct edit is then applied to this intermediate model, rather than the original model. Generalizing the procedures illustrated in this example leads to the algorithmic workflow shown in Fig.~\ref{fig:gti-resolution-framework}. In the diagram, the critical points where GTIs occur have been abbreviated as GTIPs (GTI points).

The essence of the iterative approach is to decompose a user-specified direct edit into several sequential, smaller direct edits. The benefits of doing so is that inconsistency resolution right at a GTIP can be made simple since a valid reference model is close by. By contrast, inconsistency resolution at the end of a direct edit is not under such an advantageous situation. (It should, however, be noted that, although the iterative approach applies to almost all direct modeling operations such as push-pull and face-resize, it cannot handle the face-delete operation because this operation removes boundary faces in an abrupt way and it is impossible to decompose it into smaller pieces. This can be considered as a serious limitation of the proposed method.)

\subsection{SAI detection and resolution}
\label{sec:Methodology-overview-sai}
After resolving all GTIs, the next task is to execute SAI detection and resolution modules. As already noted, SAIs take the form of over-constrained and under-constrained parts. The major task of the SAI detection module is thus to take out such parts and use them as inputs to the SAI resolution module to generate resolution options and to prioritize them for easy decision-making. However, simply picking out those parts and presenting them altogether do not provide any insights into how individual inconsistencies are formed. They should be decoupled before moving to the SAI resolution module. Decoupling is important because a constraint relevant to the resolution of one over-constrained (or under-constrained) part could be irrelevant to others. When handling one part, including any irrelevant constraints can only complicate the SAI resolution work that follows.

To accomplish the decoupling, the sizes of over-constrained parts should be minimized, and those of well-constrained parts maximized. An over-constrained part is essentially a group of constraints having dependencies. If minimized, it cannot be decomposed into smaller subparts, and then there is only one cyclical constraint dependency within it. As such, no irrelevant constraints can be included. Having only one cyclical constraint dependency in an over-constrained part also provides the following benefit: removal of any constraint can break the cyclical dependency, as shown by the resolution options in Table~\ref{tab:candidate-resolution}. In other words, the valid resolution options for resolving a minimal over-constrained part are its constraints themselves.

For under-constrained parts, a similar requirement can be stated: their sizes should be minimized. This is equivalent to maximizing every well-constrained part in the model because a model's under-constrained state can be expressed in terms of DOFs between its well-constrained parts. Once the well-constrained parts are maximized, viable resolution options are constraints that can restrict the DOFs between them (while not adding new over-constraint to the model). Any constraints that are defined within individual maximized well-constrained parts are invalid resolution options.

\begin{figure}[t]
    \centering
    \includegraphics[width=0.48\textwidth]{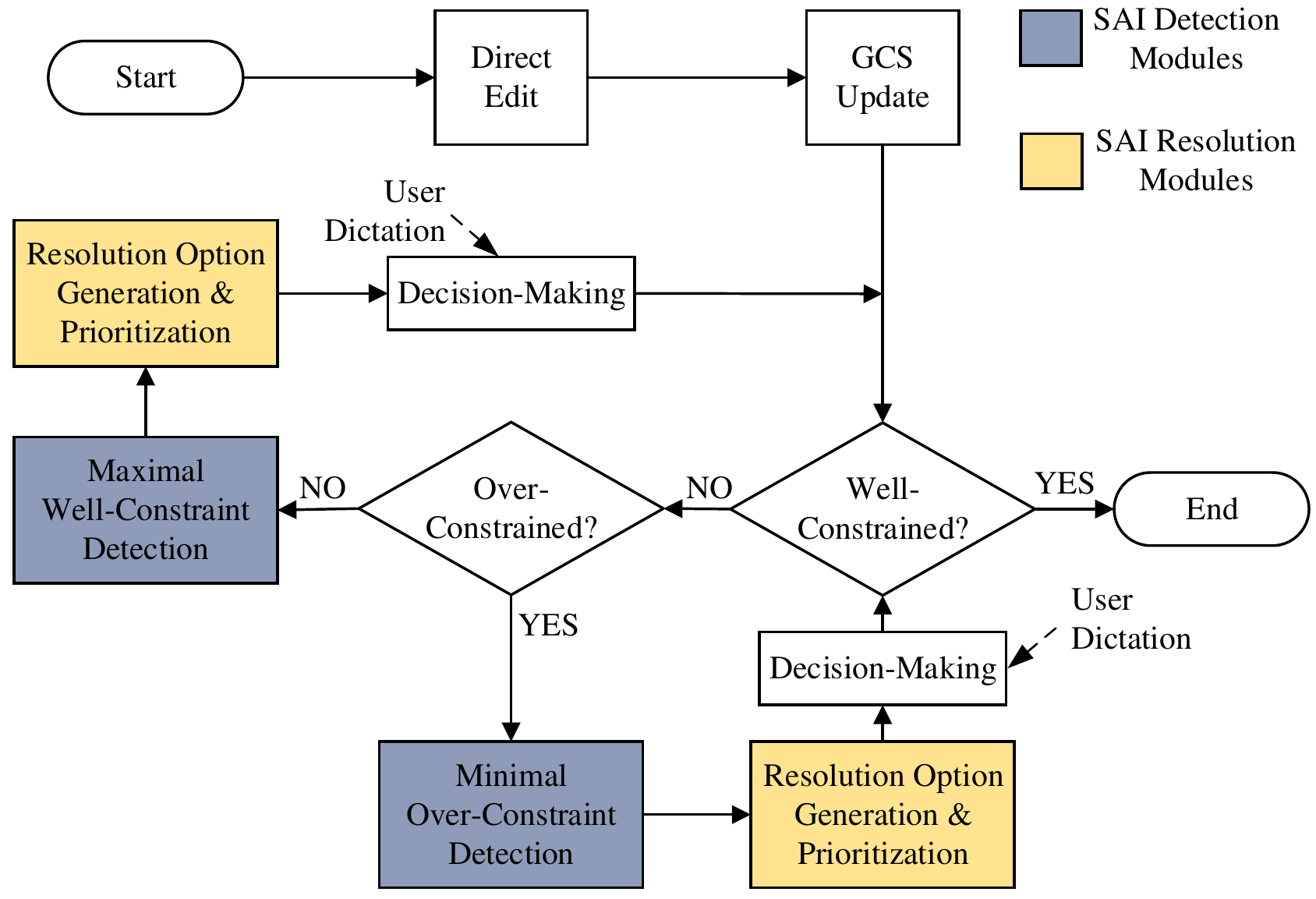}
    \caption{Schematic diagram of SAI detection and resolution.}\label{fig:sai-resolution-framework}
\end{figure}

Based on the above discussions, we outline the workflow shown in Fig.~\ref{fig:sai-resolution-framework} for SAI detection and resolution. It begins with a user-specified direct edit, then carries out GCS update according to the new model shape, then sends it to an analyzer (the modules in diamond shape) to evaluate its constraint state. If the model is still well-constrained, nothing needs to be done; otherwise, the workflow is directed to two different branches. In both branches, it first takes out information inconsistencies (the two detection modules) and then, based on the detection results, generates and prioritizes valid resolution options (the two prioritization modules), then presents the prioritized options to the user for decisions. An automatic mode is also provided in case the user chooses not to handle the inconsistencies manually. In this mode, the top options in the prioritization lists will be chosen.

\section{Implementation details of variational direct modeling}
\label{sec:Methodology-details}
Having outlined the framework, this section is devoted to the methods that can implement its constituent modules, demonstrating the practical feasibility of variational direct modeling. The discussions/formulations in the previous section has led to a breakdown of implementing the framework into solving the following four critical technical problems:
\begin{enumerate}
    \item Next GTIP detection for implementing the GTI detection module in Fig.~\ref{fig:overall-framework-flow};
    \item GTI resolution at GTIPs for implementing the GTI resolution module in Fig.~\ref{fig:overall-framework-flow}; 
    \item Minimal over-constraint and maximal well-constraint detection for implementing the SAI detection module in Fig.~\ref{fig:overall-framework-flow}; and
    \item Constraint prioritization for implementing the SAI resolution module in Fig.~\ref{fig:overall-framework-flow}.
\end{enumerate}

Next we will show how currently available algorithms can solve these critical problems, to what extent, what new improvements can be proposed to complement existing methods, and which parts still remain problematic. Basically, with existing methods and some slight improvements, a preliminary prototype can be implemented for demonstrating the framework's feasibility, but for the framework to fully function, further development is still needed, especially for the fourth technical problem. We hope that these open spaces will attract more research efforts from our community to eventually obtain a seamless integration of parametric and direct modeling. It should also be noted that the main focus and contributions of this paper lies in the theoretical foundations for parametric/direct integration and the variational direct modeling framework, as already presented in Sections \ref{sec:challenges} and \ref{sec:Methodology-overview}. The algorithms to be used to solve the above problems only represent feasible methods, not necessarily meaning the best or only ways.

\subsection{Next GTIP detection}
\label{sec:Methodology-details-GTIP-detection}
As direct modeling is a relatively new notion in the CAD domain, there has not been much research work related to GTIP detection. Existing methods \cite{lipp2014pushpull,van2010tracking,hidalgo2012computing,Zou,zou2021robust} consistently employ a heuristic strategy to check if a given modeling operation will cause GTIs and when they occur. The basic idea is: it first relates GTIPs to degenerated configurations of boundary faces (e.g., two originally intersected planes become parallel), and then checks if those generated degenerated configurations can actual happen.

Using the above strategy, Lipp et al. \cite{lipp2014pushpull} designed a set of complete heuristics to predict when GTIs will occur during a push-pull move, but only for polygonal mesh models that are composed only of planar and non-holed boundary faces. Unfortunately, such models can only make up a small portion of solid models. In our previous work \cite{Zou}, another set of heuristics have been derived to extend Lipp's work to handling models composed of linear, quadratic, and holed boundary faces (but not involving freeform surfaces). However, the developed heuristics are still restrict to local GTIs that are formed between neighboring faces. For global GTIs, e.g., one boundary face penetrating into another, both the methods presented in \cite{lipp2014pushpull,Zou} are found limited.

In parametric modeling, there are also some research studies \cite{van2010tracking,hidalgo2012computing} related to GTIP detection. Their task is to compute critical points at which the model topology changes during parametric edits. This is a little different from the problem considered here, but the notion of these critical points is conceptually similar. The methods presented in \cite{van2010tracking,hidalgo2012computing} also employ heuristics on possible degenerated face configurations to predict topology change points during a parametric modification. Nevertheless, before checking degenerated face configurations, they first identify boundary faces affected by the modeling operation to reduce checking time (to be called the culling idea).

In this work, the heuristics developed in \cite{Zou} and the idea presented in \cite{van2010tracking,hidalgo2012computing} have been combined to implement the GTIP detection module. We do not use the culling idea to accelerate GTIP detection but to handle global GTIs. Specifically, before applying the heuristics developed in \cite{Zou}, we perform boundary regeneration to see which boundary faces in the model become invalid. Boundary faces of this kind are the affected faces during the direct edit. Clearly, affected faces so obtained comprise a superset of the boundary faces related to global GTIs (if any) during a direct edit.

\begin{figure}[t]
    \centering
    \includegraphics[width=0.48\textwidth]{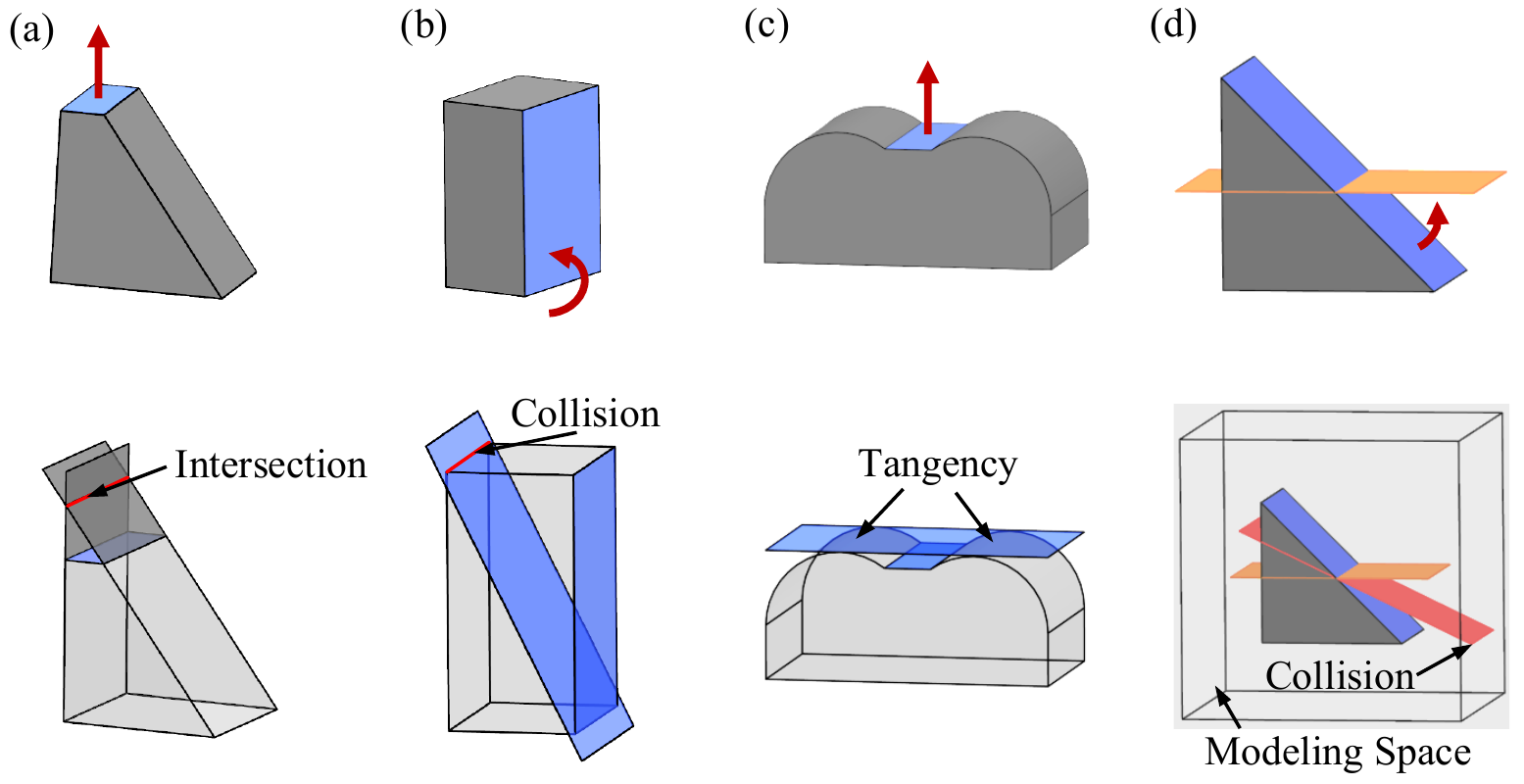}
    \caption{Illustration of degenerated configurations in GTIP detection (blue face: push-pulled face; curvy arrows: rotational push-pull; straight arrows: translational push-pull) \cite{Zou}.}\label{fig:degenerated-configurations-gti}
\end{figure}

With affected boundary faces in place, we directly apply the heuristics from \cite{Zou} to them to generate an exhaustive list of degenerated configurations, then check if any of these candidates can actual happen for a specific direct edit, resulting in a reduced set of candidates. Finally, we pick the closest remaining candidate as the desired next GTIP. The above procedures have been summarized in Algorithm~\ref{alg:alg1}. This algorithm will not miss any GTIPs because the heuristics from \cite{Zou} were developed in a systematic way. A brief version of the derivation is as follows. As already noted in Section~\ref{sec:challenges}, there are two types of sources for GTIs: inserting new connections and losing old connections. Exhaustive combinations of these two sources and surface types give three possible ways GTIs could occur, and they can be translated to the following four degenerated configurations in total: surface-surface intersection, surface-face collision, surface-surface tangency, and face-workspace collisions, as illustrated in Fig.~\ref{fig:degenerated-configurations-gti}. Here we have distinguished faces that are short for boundary faces, and surfaces that underlie boundary faces.

A practical note should be noted here. Doing an exhaustive checking against all generic degenerated configurations is time-consuming for general cases, but this is acceptable in the context of direct modeling because direct edits are local modifications and the actual number of boundary faces involved in the checking is usually very small.

\begin{algorithm}[t]
    \includegraphics[width=0.45\textwidth]{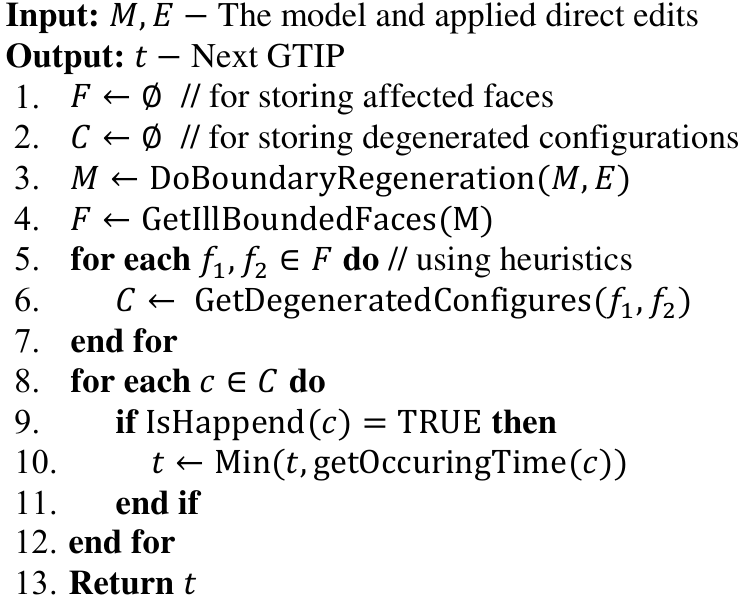}
    \hspace*{\fill}
    \caption{Next GTIP Detection}
    \label{alg:alg1}	
\end{algorithm}

\subsection{GTI resolution at GTIPs}
\label{sec:Methodology-details-GTIP-Resolution}
\begin{figure}[htbp]
    \centering
    \includegraphics[width=0.48\textwidth]{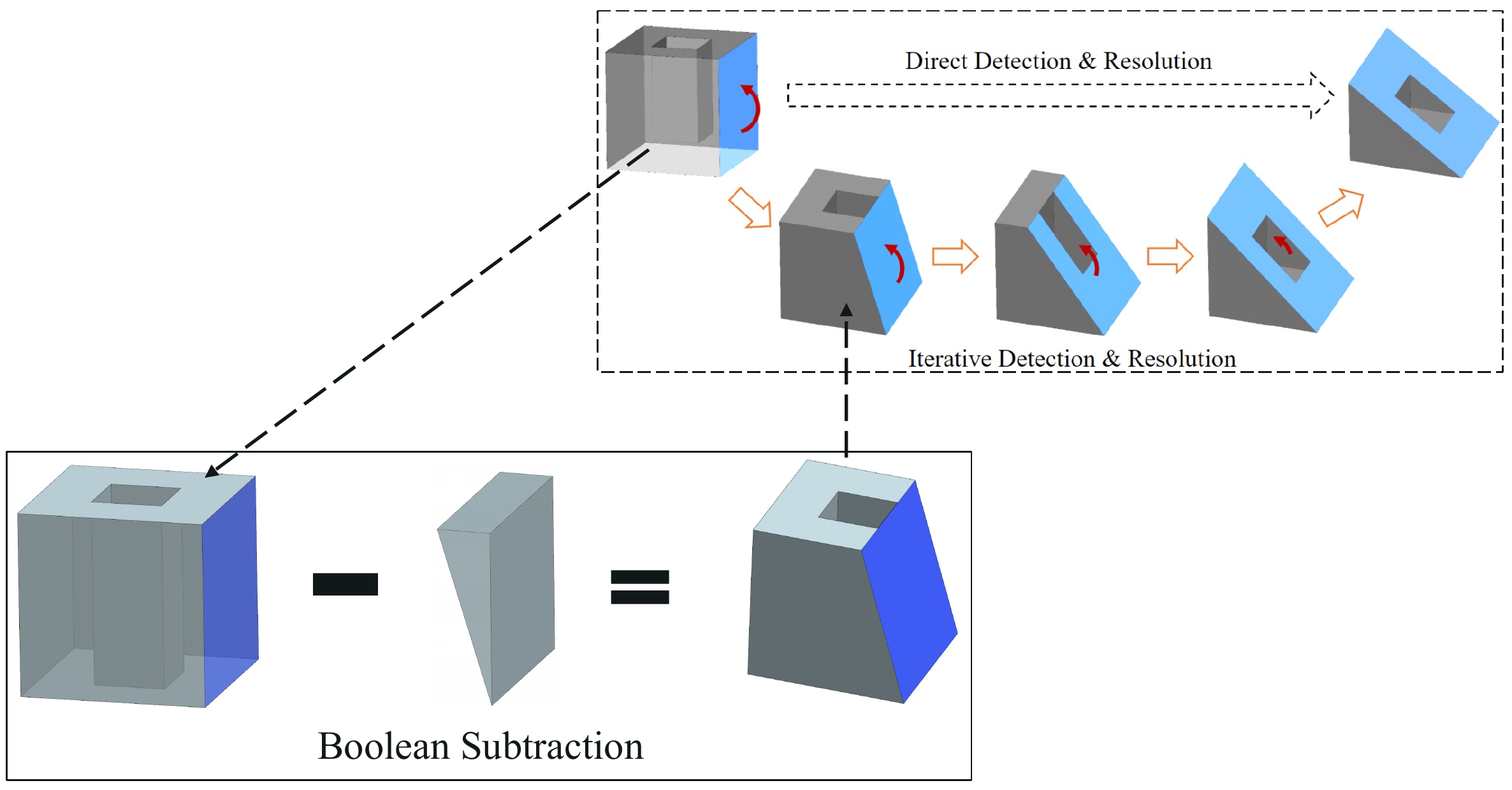}
    \caption{Illustrations of Boolean-based GTI resolution (the upright subfigure is the same as that in Fig.~\ref{fig:progressive-gti-framework}).}\label{fig:boolean-gti-resolution}
\end{figure}

GTI resolution is to modify the before-GTIP topology to accommodate the after-GTIP geometry while ensuring valid modeling results and continuous model variations. This task is also under the same situation that there is little research work \cite{lipp2014pushpull,Zou,zou2021robust} for it. Lipp et al. \cite{lipp2014pushpull} proposed to directly modify model topology to resolve any detected GTIs through, again, a set of heuristics. Robustness of those heuristics for polygonal mesh models have been demonstrated, but how to extend them to handling models composed of linear and quadratic surfaces remains unknown. Different from their direct topology modification method, our previous work \cite{Zou,zou2021robust} has developed an indirect approach, which transforms the task here to Boolean operations on the model volume. Fig.~\ref{fig:boolean-gti-resolution} illustrates the transformation, based on the first GTIP in Fig.~\ref{fig:progressive-gti-framework}. The desired intermediate model for this GTIP can be directly obtained through subtracting the wedge-shaped model from the original model. Use of Boolean operations instead of direct topology modification can provide the important advantage that validity of the resulting model is automatically guaranteed \mbox{\cite{requicha1985boolean}}. Also, there is no need to get down to the model's low-level topological and geometric data, which is otherwise error-prone. In this work, this approach has been adopted to implement the GTI resolution module.

\begin{figure*}[htbp]
    \centering
    \includegraphics[width=0.6\textwidth]{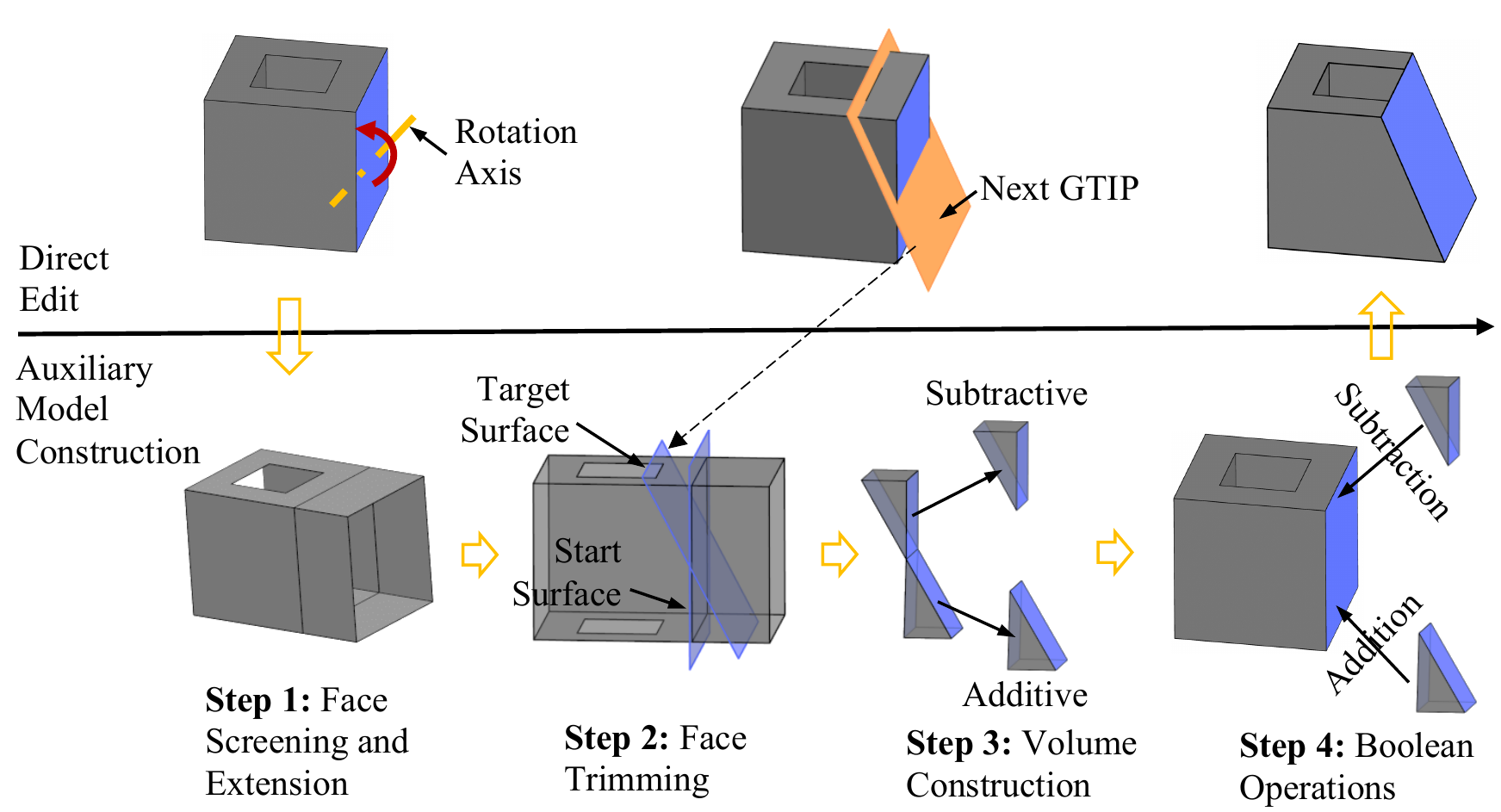}
    \caption{Illustration of the procedures used for auxiliary model construction.}\label{fig:boolean-gti-auxiliary}
\end{figure*}

It is also easy to attain continuous model variations using the Boolean-based method. What we need to do is to add continuity constraints to the construction of volumes to be subtracted from or added to the original model. These volumes are to be called auxiliary models. The specific continuity constraint is \cite{Zou}: restricting the auxiliary model to the volume swept by the edited surface from the current point to the next GTIP, as well as bounded by its neighboring surfaces. A four-step construction algorithm has been developed for this purpose, as illustrated in Fig.~\ref{fig:boolean-gti-auxiliary} and summarized below:
\begin{itemize}
    \itemsep0em
    \item The first step is to pick out relevant neighboring boundary faces and extend them to cover the whole sweeping range of the direct edit.
    \item The second and third steps work together to get the volume bounded by the neighboring boundary faces generated in the first step and the directly edited boundary faces. The volume generated in this way has two possible types: if the direct edit adds material to the model, the volume is additive; if it removes material, the volume is subtractive, as shown in Fig.~\ref{fig:boolean-gti-auxiliary}.
    \item The last step carries out the actual Boolean operations between the original model and the auxiliary models generated by the previous steps.
\end{itemize} 

It should be noted here that the above steps only represent a brief introduction to the algorithm. It clearly involves more technical details and singular cases than the simple case illustrated in Fig.~\ref{fig:boolean-gti-auxiliary}. For more details of how to handle such issues, please refer to \cite{Zou,zou2021robust} for a complete version of the algorithm.

\subsection{Minimal over-constraint and maximal well-constraint detection}
\label{sec:Methodology-details-sai-detection}
The detection task here includes the evaluation of the model's constraint state and, if not well-constrained, the extraction of minimal over-constrained parts and maximal well-constrained parts in the model. This is a topic extensively studied in the field of geometric constraint solving, see \cite{bettig2011geometric,hu2021geometric} for a thorough review. The proposed methods may be classified into the following four categories.

\textbf{Solving-Based} \quad This category of methods analyze a model's constraint state by directly solving it through numerical methods (e.g., homotopy continuation) or symbolic methods (e.g., Grobner bases) \cite{bettig2011geometric}. These methods are rarely used in today's CAD systems because they are very time-consuming.

\textbf{Logic-Based} \quad This category of methods apply the concept of axiomatization to GCS. Their way of working relies on a set of geometric theorems and derivation rules \cite{dufourd1998geometric}. If a model GCS can be logically derived from the theorems and rules, it is well-constrained; if there are extra constraints, it is over-constrained; otherwise, it is under-constrained. Despite the mathematical elegance, they suffer from the issue of having an exhaustive set of geometric theorems and derivation rules so that commonly used geometric constraints can be included.

\textbf{Graph-Based} \quad This category is the most popular in the literature. The basic idea is first converting a model GCS to a graph, then analyzing this graph to obtain constraint state information instead of using the original GCS. From this idea, two lines of development have been established. The first line tries to recognize subgraph patterns that correspond 
to known shapes from a given constraint graph. It is pioneered by Owen \cite{owen1991algebraic} and much
improved in \cite{bouma1995geometric,fudos1997graph,gao2002solving} in the size of the pattern library. The second line compares DOFs of the model geometry with degrees of restriction of the model GCS. It was first proposed by Bardord \cite{barford1987graphical} and Serrano \cite{serrano1987constraint}, and detailed in \cite{ait2014reduction,latham1996connectivity,hoffmann1997finding}. In 2001, these two lines were unified under a framework proposed by Hoffmann et al. \cite{hoffman2001decomposition}. Ever since, there has still been some good progress, e.g., those discussed in \cite{hu2021geometric}, but the foundations remain unchanged. Graph-based methods have seen wide applications by industry. They are, however, unable to handle GCS having constraint dependencies (except for the simplest structural dependencies) \cite{hoffmann2004making,michelucci2006geometric}. This is because, once the model GCS is converted to a graph, only its combinatorial information is retained, and all geometric information is discarded. This limitation makes graph-based methods inapplicable to the SAI detection problem considered in this work. (It should be noted that graph-based methods have a very large body of papers, and the above discussion only covers some important research studies due to page length limit.)

\textbf{Perturbation-Based} \quad To overcome the limitation of graph-based methods, Michelucci et al. \cite{michelucci2006geometric} proposed the witness configuration method (WCM). This method examines how constraint equations behave under infinitesimal perturbations made to the constraint equation variables, and the behavior is described by the associated Jacobian matrix of the constraint equations. Different constraint states have different perturbation behavior \cite{thierry2011extensions}. Despite its effectiveness in handling constraint dependencies, the presented algorithms have difficulties in minimizing over-constraint and maximizing well-constraint due to the use of greedy algorithms. For this reason, a series of WCM-based algorithms have been proposed in our prevous work \cite{zou2019variational,zou2020decision} to automatically extract the minimal over-constrained parts and maximal well-constrained parts in a model GCS.

Considering that SAIs often have constraint dependencies, WCM has been adopted to accomplish the constraint state evaluation task when implementing the SAI detection module. For the other task of minimal over-constraint and maximal well-constraint extraction, the algorithms developed in \cite{zou2019variational,zou2020decision} have been used. A brief introduction to the adopted methods is given below.

Using WCM, a model's constraint state can be characterized by its constraint equations' Jacobian matrix \cite{michelucci2006geometric}. If the Jacobian matrix has linearly dependent rows, the GCS is over-constrained. The null space of the Jacobian matrix indicates the model's DOFs and, if it has more DOFs than the six canonical rigid-body transformations, the model is under-constrained.\cite{thierry2011extensions,zou2019variational}. From linear algebra, we know that dependent rows in the Jacobian matrix $J$ are reflected by solutions to the linear system $J^T \cdot x = 0$. A vector $x$ in the null space of $J^T$ represents an over-constrained part, and the nonzero elements of $x$ indicate the constraints involved in this part. To minimize this part's size is thus equivalent to requiring that the vector $x$ has the minimum number of nonzero elements. Then, we can attain the minimized over-constrained parts by solving the following optimization problem:
\begin{equation*}
    \min _ { x } \| x \| _ { 0 } \quad \text { s.t. }\quad  J ^ { T } x = 0 , \ x \neq 0
    \label{eq:minimal-group-modeling}
\end{equation*}
where $\| \cdot \| _ { 0 }$ is the $\ell _ { 0 }$ norm whose mathematical meaning is to count non-zero elements in a vector. As such, the problem of minimal over-constraint detection is transformed into a sparse recovery (a.k.a. compressive sensing) problem, which is a well-researched problem in the image processing \cite{fornasier2015compressive} and can be solved numerically using the relaxation method described in \cite{Osher2014}.  Using a series of mathematical manipulations, the problem of maximal well-constraint detection can also be transformed into a similar sparse recovery problem, refer to Section 4.3 of \cite{zou2020decision} for a detailed derivation.

\subsection{SAI resolution option prioritization}
\label{sec:Methodology-details-sai-resolution}
According to the workflow outlined in Fig.~\ref{fig:sai-resolution-framework}, the next step after getting the minimal over-constrained and maximal well-constrained parts is to generate resolution options and then prioritize them. Generating valid resolution options is easy, given that those parts have been minimized/maximized (as already discussed in Section \ref{sec:Methodology-overview-sai}). The remaining critical task is thus to prioritize the generated resolution options.

Resolution options take the form of geometric constraints to be removed from or added to the model. Prioritizing them is to put them in a certain order, which is determined by a binary comparison operation that accepts two constraints as arguments and determines which of them should occur first. Different from the topic of geometric constraint solving, there has been much less research work on geometric constraint prioritization. The presented methods may be roughly classified into qualitative methods and quantitative ones. Qualitative methods prioritizes constraints based on a set of heuristics on their type \cite{murugappan2009towards,mills2001estimate,martinez2005constraint,zou2007constraint}. Type-based prioritization can be effective in some application scenarios because certain types of constraints could carry more engineering knowledge and occur more frequently than others. However, these methods cannot further prioritize geometric constraints of same type. This is where quantitative methods can help. A deviation-based measure has been proposed in the literature to prioritize geometric constraints \cite{langbein2004choosing,li2011globfit}. A similar idea has been used in this work to implement the SAI resolution module.

In this work, the qualitative and quantitative strategies have been combined to accomplish the task of constraint prioritization. This leads to a two-level comparison scheme consisting of a rough comparison and a fine comparison. The rough comparison is responsible for keeping constraints whose type may carry more design intent. For example, parallelism is likely to carry more design intent than a general angle constraint. The specific type precedence used in this work can be found in Table~\ref{tab:rough-prioritization}, based on the experimental data from \cite{zou2007constraint,langbein2004choosing}.

\begin{table}[htbp]
    \centering
    \caption{Rough prioritization based on constraint types.} 
    \includegraphics[width=0.48\textwidth]{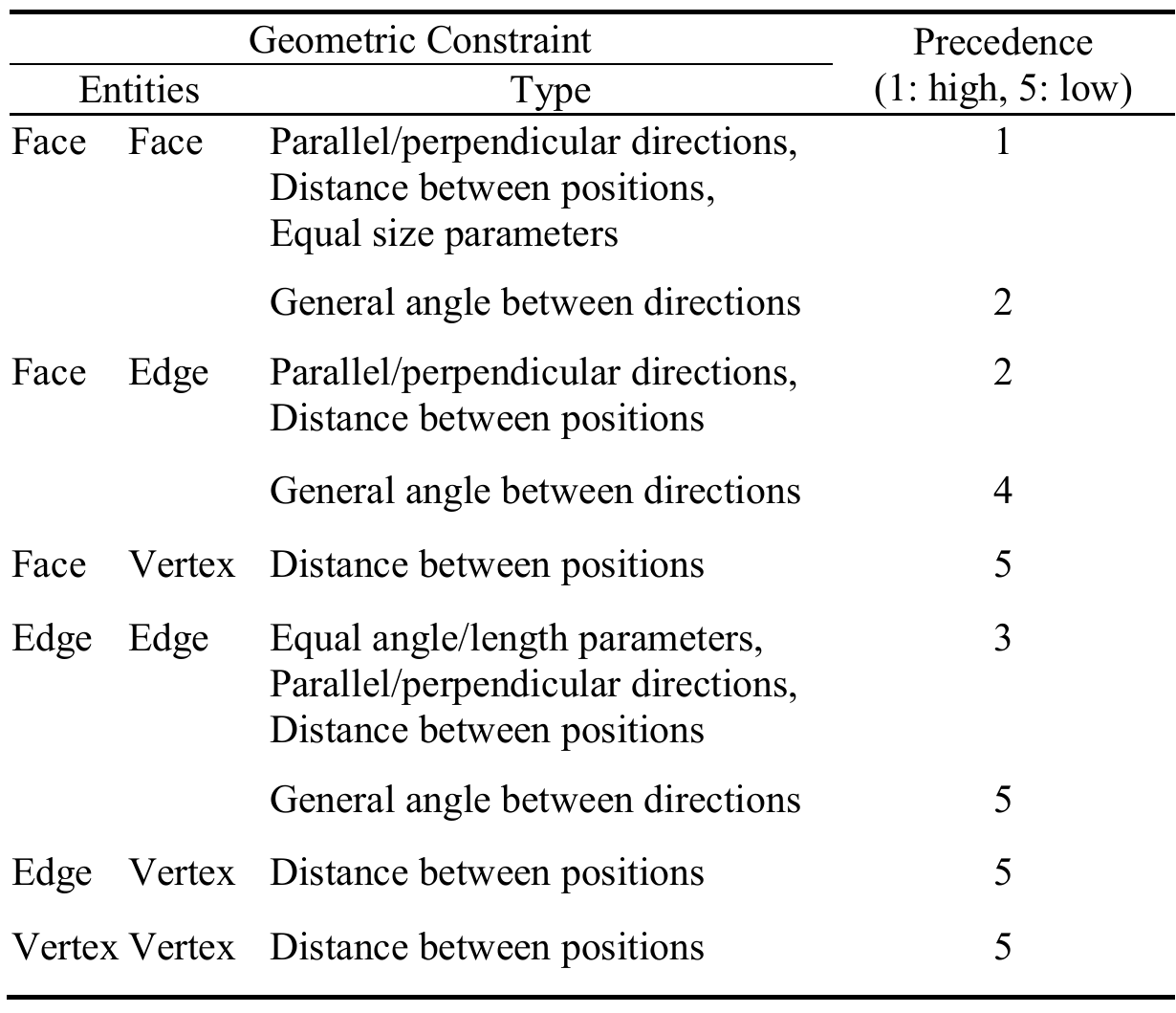}
    \label{tab:rough-prioritization}
    \vspace{-4mm}
\end{table}

The fine comparison takes care of geometric constraints of same type. We compare them based on the concept of sensitivity. Parameter changes made to different constraints often lead to varied model shape changes, which can be mathematically characterized by the notion of sensitivity (i.e., the rate of the model shape change with respect to parameter changes). If a constraint has a high sensitivity value, a small parameter change made to it will result in a large change on the model shape, which may lead to a dramatic, unpredictable shape change. A model under such a situation is said to have a poor constraining scheme \cite{hillyard1978analysis}. We thus use sensitivity to quantify a constraint's impact on the model shape. With this choice, it is expected that resolution options leading to a model with the least change rate will be chosen, then the least model variations could be expected for later parametric edits.

It should be noted that the above constraint prioritization method still employs heuristics. Therefore, it may have the generalization issue. Currently, we mitigate this issue by incorporating user guidence (if necessary), as shown in Fig.~\ref{fig:sai-resolution-framework}. However, prompting the user to choose an appropriate result may affect his/her workflow. In this regard, a constraint prioritization method that can work in an automatic and intelligent way is much desired. This is where the state of the art is not enough, and new developments are needed.

\begin{figure*}[!htb]
    \centering
    \includegraphics[width=0.8\textwidth]{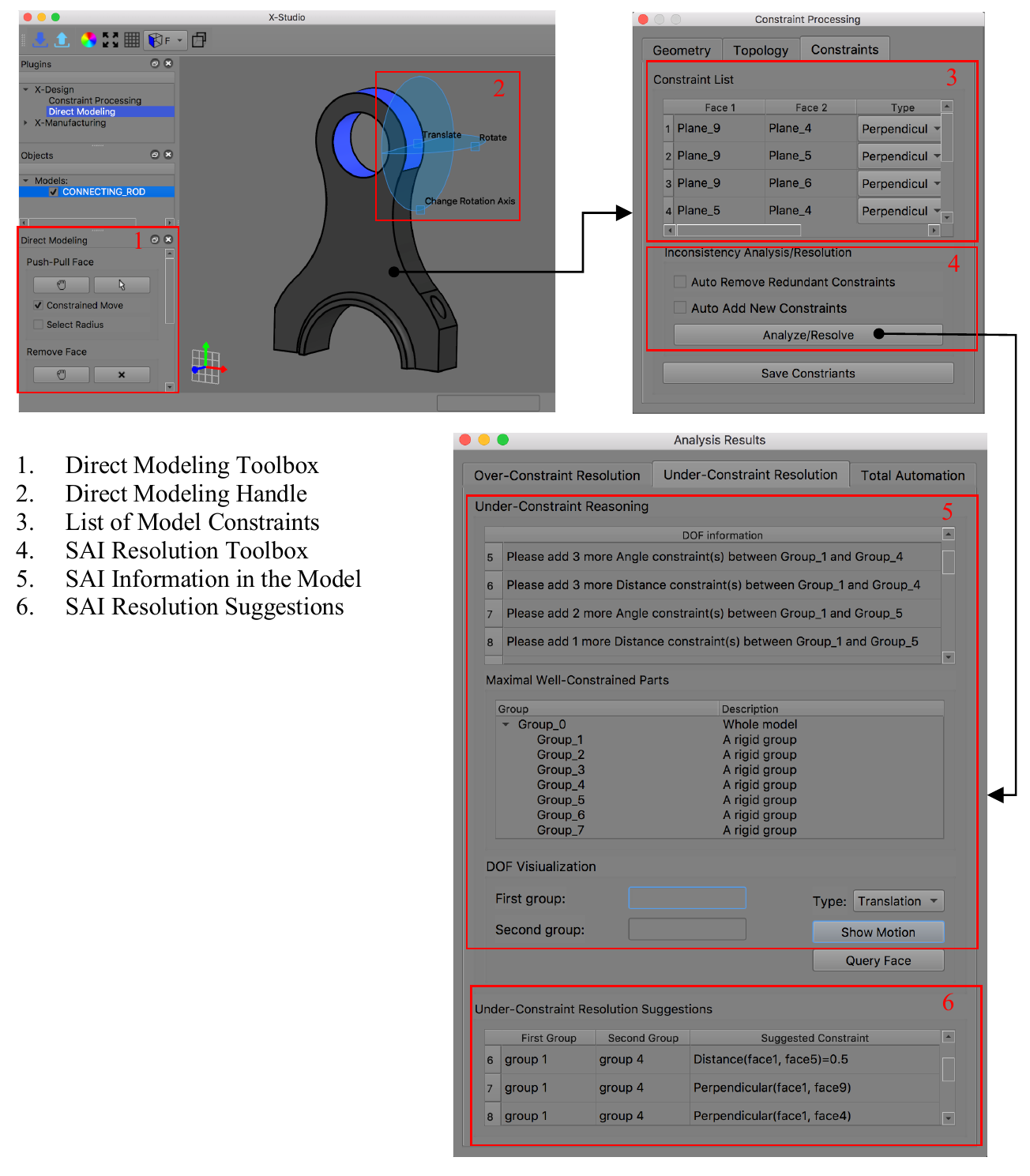}
    \caption{Graphical user interface of the prototype modeler.}\label{fig:sci-gui}
\end{figure*}

\section{Modeling examples}
\label{sec:results}

\subsection{Prototype implementation}
\rev{A preliminary prototype modeler has been developed to show effectiveness of the proposed framework, . The modeler's GUI is shown in Fig.~\ref{fig:sci-gui}, which implements the interface module in Fig.~\ref{fig:overall-framework-flow}. It was developed using C++ and QT (version 5.7), taking the open-sourced geometry processing and rendering framework OpenFlipper as a reference. The information maintenance module in Fig.~\ref{fig:overall-framework-flow} was implemented on top of Open CASCADE (version 7.0), an open source CAD modeling kernel.}
{To show effectiveness of the proposed framework, a preliminary prototype modeler has been developed using C++ in an Apple Macintosh environment (2.4 GHz Intel Core i5 with 8G memory). The software's architecture is similar to the opensource geometry processing and rendering framework OpenFlipper (version 3.0). The modeler's GUI is shown in Fig.~\ref{fig:sci-gui}, which implements the interface module of the framework outlined in Fig.~\ref{fig:overall-framework-flow}. It was implemented using the QT library (version 5.7). The other four modules of the framework, i.e., GTI detection, GTI resolution, SAI detection, and SAI resolution, were developed on top of the geometric modeling kernel Open CASCADE (version 7.0). All the numeric solving/optimization tasks were carried out using the C++ library Eigen (version 3.2.9) and MATLAB.}

To edit a model, the user clicks buttons in the direct modeling toolbox (red box 1), then a graphical model manipulation handle pops up (red box 2) to allow the user to make interactive changes. To view the model's geometric constraints, the user clicks the constraint processing button in the left sidebar, then the middle panel shown in Fig.~\ref{fig:sci-gui} pops up, with all constraints listed in red box 3. To see if there are any SAIs in the model, the user clicks the Analyze/Resolve button in red box 4, then the computer presents all information necessary to resolve them (red boxes 5 and 6), including DOFs, constraint dependencies, and resolution suggestions. If the user wants the computer to take care of all work, the two Auto options in red box 4 should be checked.

\begin{figure*}[htbp]
    \centering
    \includegraphics[width=0.7\textwidth]{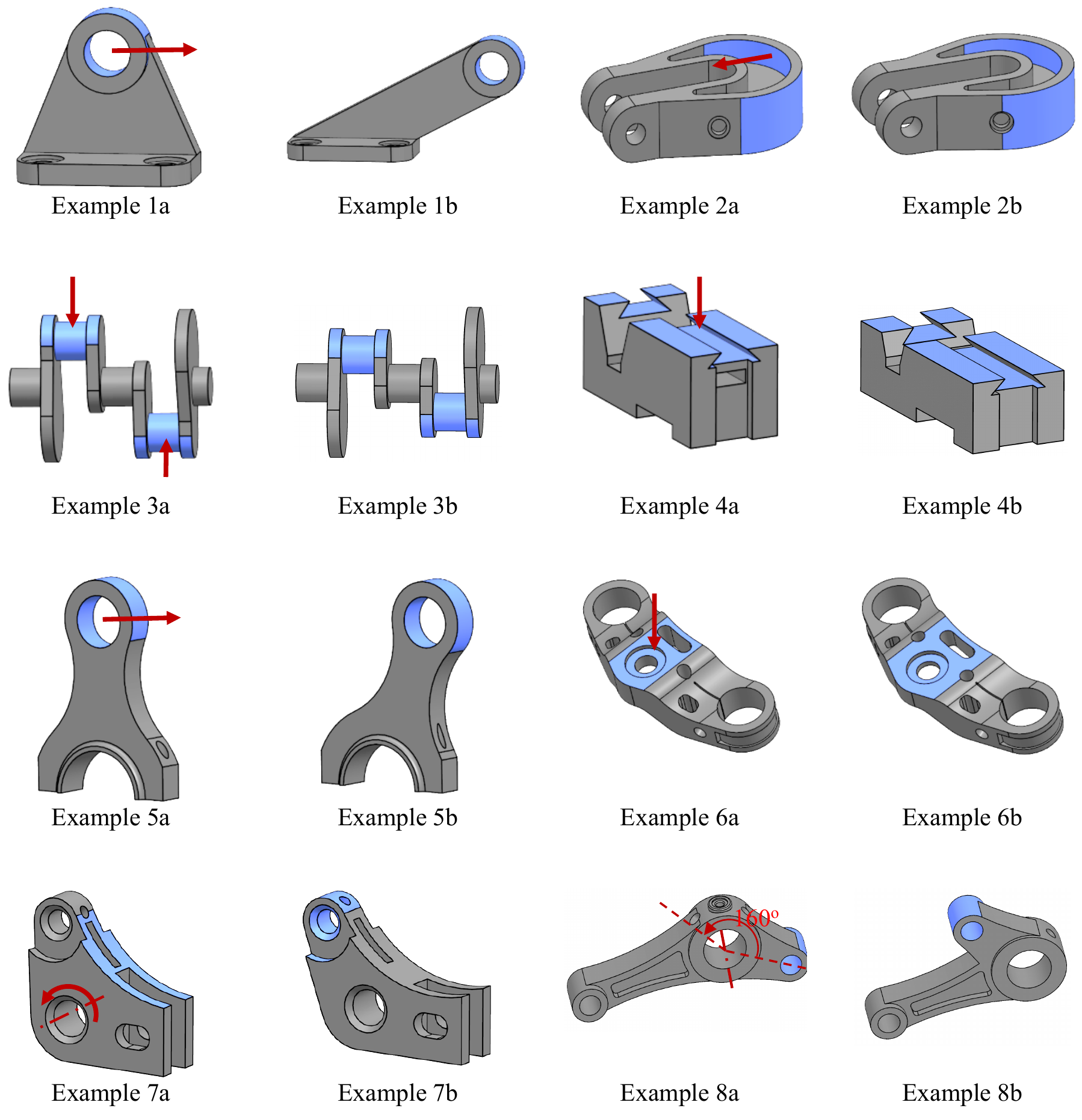}
    \caption{Examples of GTI detection and resolution (a: the original model and the applied direct edit; b: the resulting model).}\label{fig:result-all-cases-gti}
\end{figure*}

\subsection{Examples and comparisons}
\begin{figure*}[htbp]
    \centering
    \includegraphics[width=0.6\textwidth]{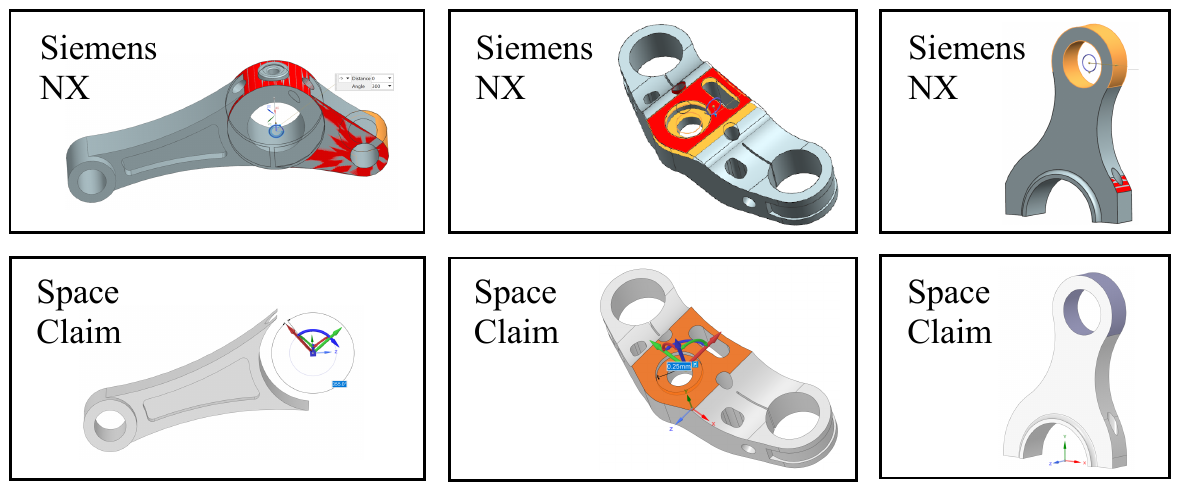}
    \caption{Sample failure examples given by the two most advanced direct modelers, Siemens NX and SpaceClaim, for the examples in Fig.~\ref{fig:result-all-cases-gti}.}\label{fig:result-all-cases-gti-comparisons}
\end{figure*}

Based on the prototype modeler, the usefulness of the proposed detection and resolution mechanism (i.e., those modules in Figs.~\ref{fig:gti-resolution-framework} and \ref{fig:sai-resolution-framework}) has been tested with examples taken from both real and simulated data. Fig.~\ref{fig:result-all-cases-gti} shows some of the test examples. The blue faces are the edited faces, the red arrows depict push-pull directions, and the dashed red lines are rotation axes. For example, in Example 1a, the blue faces were rotated counterclockwise by angle of $160$ degrees. \rev{Fig.~\ref{fig:result-all-cases-gti-comparisons}a}{Table~\ref{tab:result-all-cases-gti-comparisons}} summarizes the comparison results with leading CAD software, based on the examples shown in Fig.~\ref{fig:result-all-cases-gti}. Some of the failure cases of Siemens NX and Ansys SpaceClaim are also given in Fig.~\ref{fig:result-all-cases-gti-comparisons}. Usually, NX signals model update failures by coloring relevant faces in red, and SpaceClaim does so by either coloring relevant faces in orange (e.g., the middle one) or giving a random, yet wrong shape (e.g., the left one).

\newcolumntype{Y}{>{\centering\arraybackslash}X}
\begin{table*}[htbp]
    \caption{Comparisons with leading CAD software and their failure/success stats.} 
    \centering
        \begin{tabularx}{0.75\textwidth}{l c*{7}{Y} c}
        \toprule
        \multirow{2}{*}{}   &  \multicolumn{8}{c}{Test Models (Refer to Fig.~\ref{fig:result-all-cases-gti} for Numbering)}  & \multirow{2}{6em}{\centering Overall Success Ratio}  \\
        \cmidrule(lr){2-9}
                            & 1 & 2 & 3 & 4 & 5 & 6 & 7 & 8 &   \\
        \midrule
        Siemens NX           & $\checkmark$ & $\checkmark$ & $\times$ & $\times$ & $\times$ & $\times$ & $\checkmark$ & $\times$ &  3/8 \\
        SpaceClaim          & $\checkmark$ & $\times$ & $\times$ & $\times$ & $\times$ & $\times$ & $\checkmark$ & $\times$ & 2/8  \\
        PTC Creo            & $\checkmark$ & $\checkmark$ & $\times$ & $\times$ & $\times$ & $\times$ & $\times$ & $\times$ &  2/8 \\
        Autodesk Inventor   & $\checkmark$ & $\checkmark$ & $\times$ & $\times$ & $\times$ & $\times$ & $\times$ & $\times$ & 2/8  \\
        SolidWorks          & $\checkmark$ & $\times$ & $\checkmark$ & $\times$ & $\times$ & $\times$ & $\times$ & $\times$ & 2/8  \\
        The Proposed Method & $\checkmark$ & $\checkmark$ & $\checkmark$ & $\checkmark$ & $\checkmark$ & $\checkmark$ & $\checkmark$ & $\checkmark$ & 8/8 \\
        \bottomrule
        \end{tabularx}
    \label{tab:result-all-cases-gti-comparisons}
\end{table*}

\begin{figure*}[htbp]
    \centering
    \includegraphics[width=0.7\textwidth]{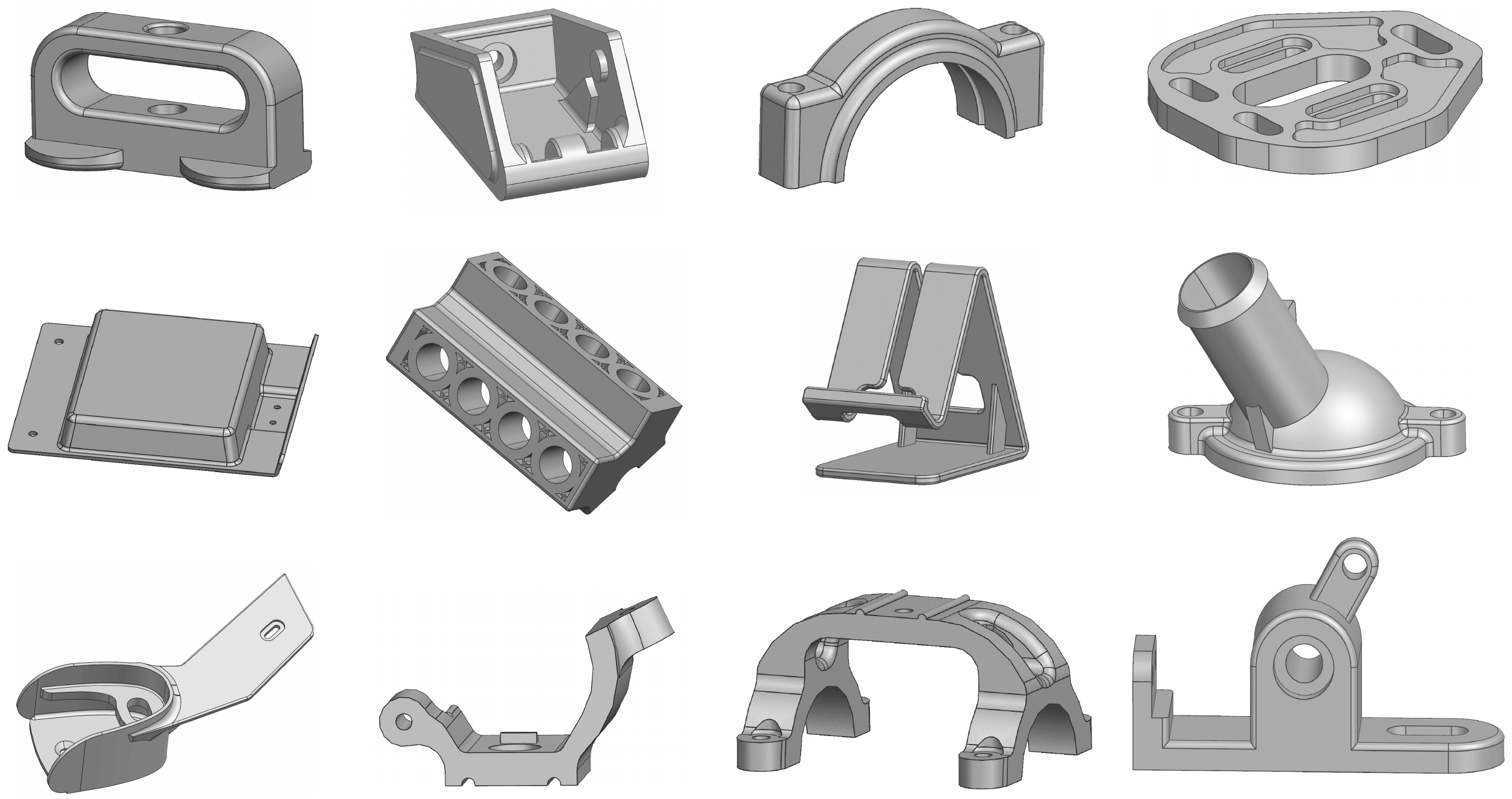}
    \caption{Models, together \rev{}{with} those in Fig.~\ref{fig:result-all-cases-gti}, used to conduct quantitative comparison experiments for Siemens NX, Ansys SpaceClaim, and ours.}\label{fig:result-quant-analysis-gti}
\end{figure*}

A more quantitative comparison between Siemens NX, Ansys SpaceClaim, and ours has also been carried out on a larger set of 20 CAD models, including those already presented in Fig.~\ref{fig:result-all-cases-gti} and an additional set of the 12 CAD models shown in Fig.~\ref{fig:result-quant-analysis-gti}. The testing was carried out by applying 10 random direct edits to each test model and then measuring the methods' success ratios. For the $200 = 20 \times 10$ random direct edits, half of them were purposefully set to not causing any topology changes on the models being edited. The final success ratios are as follows. For the 100 direct edits that not cause topology changes, the success ratios of the three methods/tools are very close: Siemens NX (94\%), Ansys SpaceClaim (96\%), and ours (96\%). For the rest 100 direct edits that involve topology changes, big difference has been observed: Siemens NX (76\%), Ansys SpaceClaim (84\%), and ours (93\%). This confirms the proposed method's robustness in handling information inconsistencies.
  
\begin{figure*}[htbp]
    \centering
    \includegraphics[width=0.7\textwidth]{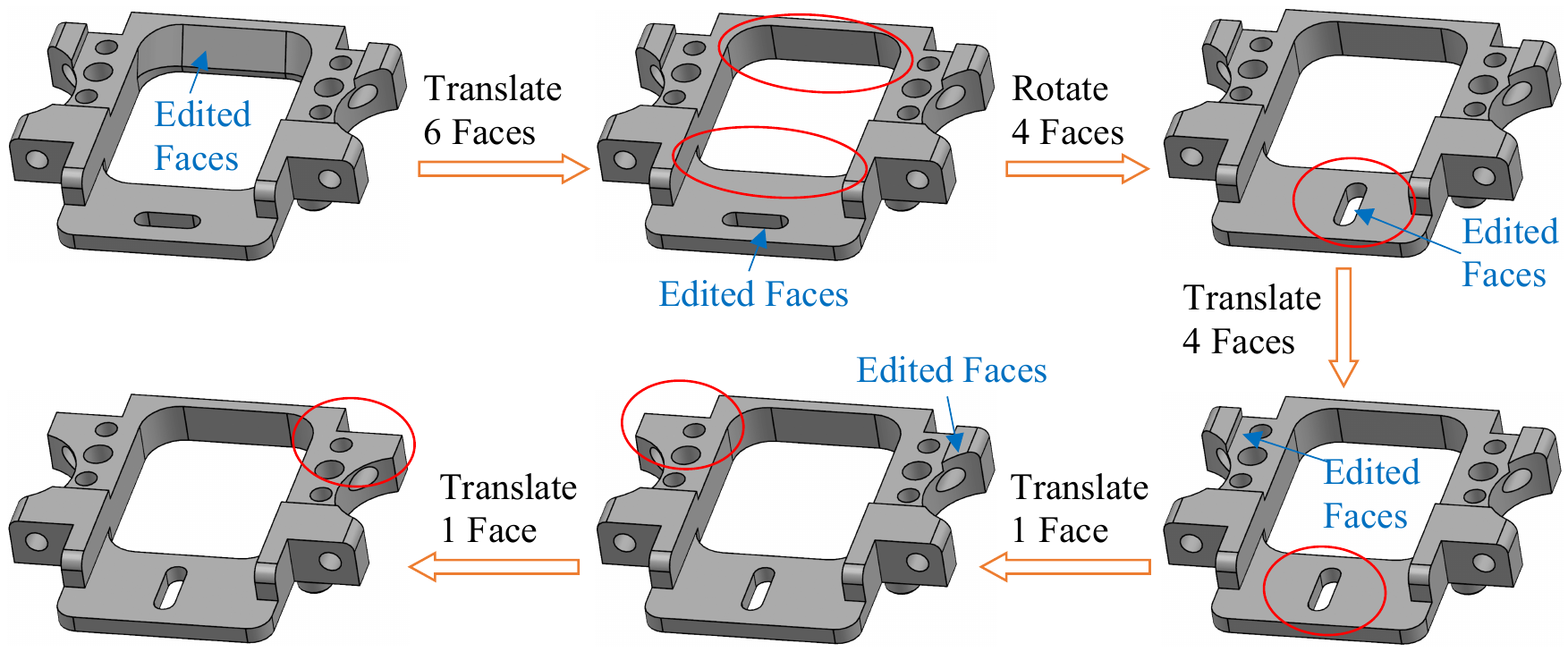}
    \caption{A series of direct edits on a gear box mount model (circles indicate changed parts) \cite{Zou}.}\label{fig:result-gti-single}
\end{figure*}

Fig.~\ref{fig:result-gti-single} shows one example where all the modules (i.e., the previously presented GTI detection and GTI resolution modules) are assembled to work together. It involves a gear box mount part model (obtained from the GrabCAD part library https://grabcad.com/library). The applied direct edits are quite comprehensive, including single-face edit, multiple-face edits, the translational edit type, and the rotational edit type. In each subfigure, the blue text and arrow indicate the faces to be changed by the direct edit that follows, and those circled out by red ellipsed show the editing results. From all the examples and comparisons in Figs.~\ref{fig:result-all-cases-gti}, \ref{fig:result-all-cases-gti-comparisons} and \ref{fig:result-gti-single}, the proposed method is seen to improve significantly the robustness of direct modeling.

\begin{figure*}[htbp]
    \centering
    \includegraphics[width=0.75\textwidth]{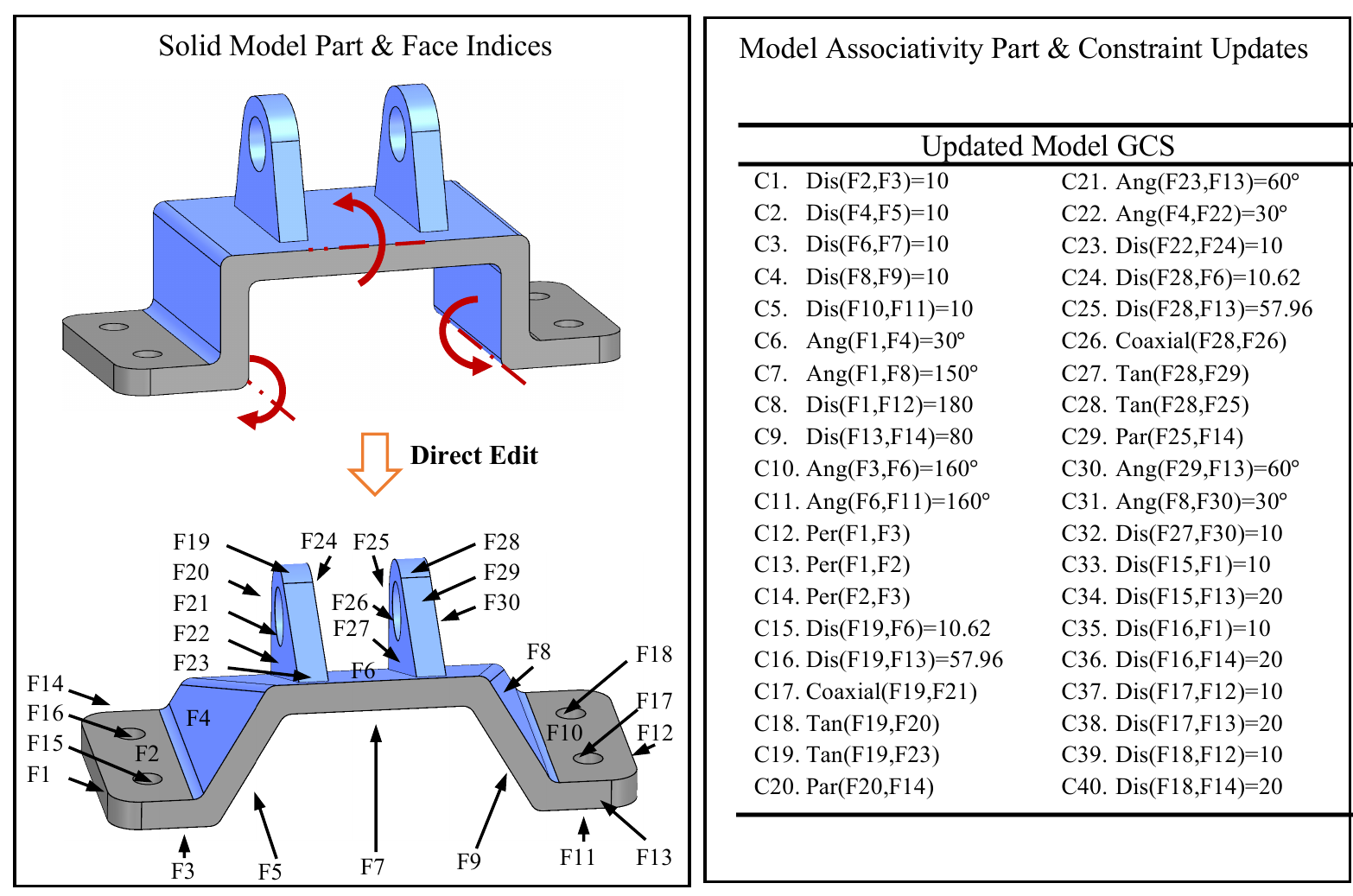}
    \caption{Left: engine bracket model and a direct edit applied to it; right: updated model GCS (Dis: Distance, Ang: Angle, Per: Perpendicular, Par: Parallel, Tan: Tangent).}\label{fig:result-sai-bracket}
\end{figure*}

\begin{figure*}[htbp]
    \centering
    \includegraphics[width=0.6\textwidth]{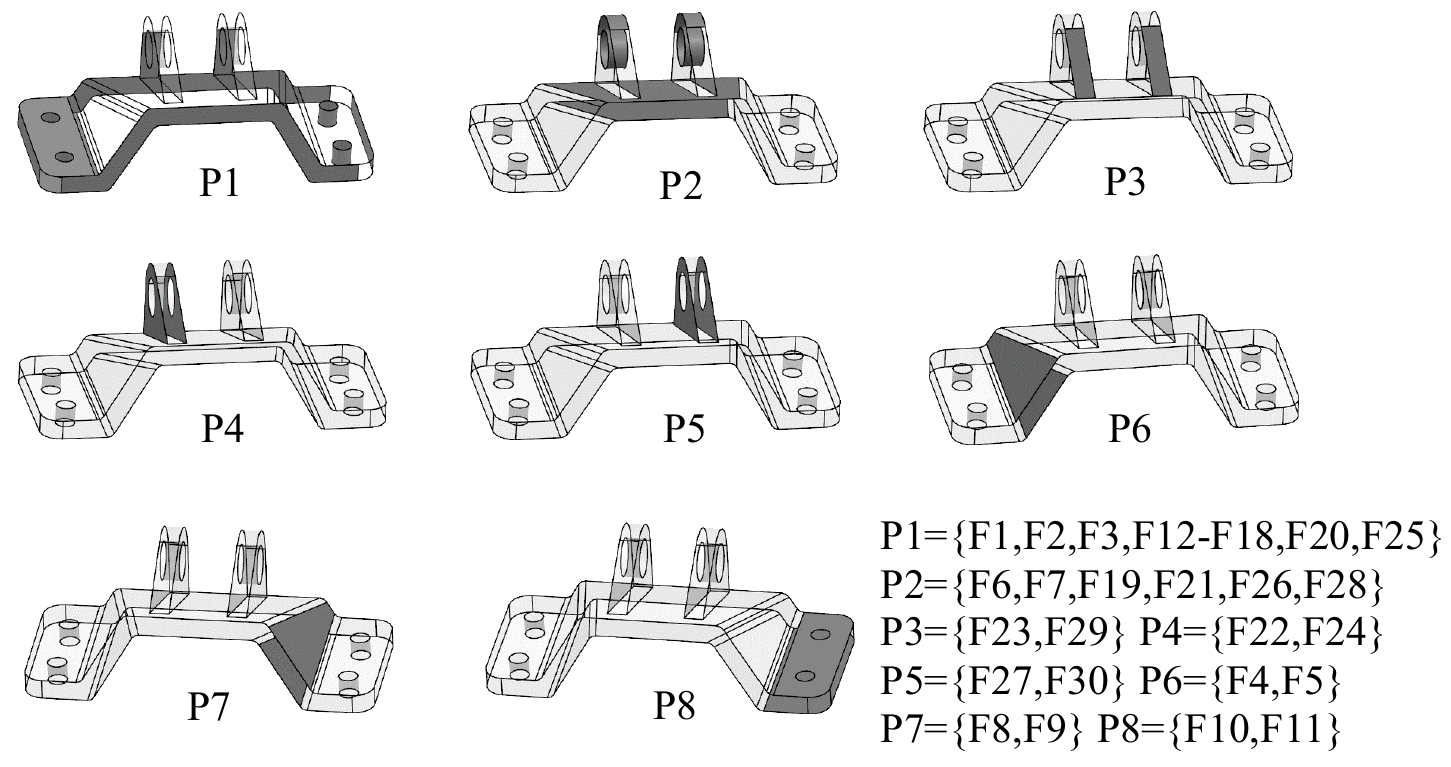}
    \caption{Detected maximal well-constrained parts for the engine bracket model (dark faces: the belonging faces of individual parts; transparent faces: the non-belonging faces; face lists: the belonging faces' indices as in Fig.~\ref{fig:result-sai-bracket})}.\label{fig:result-sai-bracket-resuelt}
\end{figure*}

\begin{figure}[htbp]
    \centering
    \includegraphics[width=0.48\textwidth]{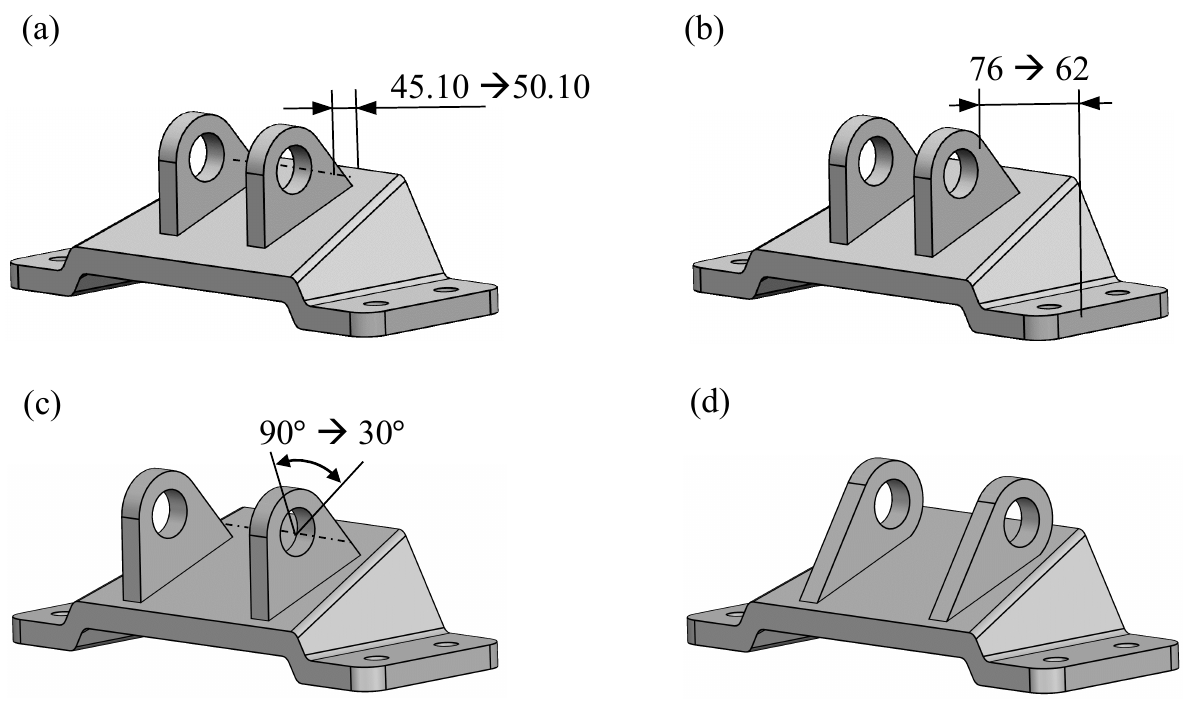}
    \caption{A series of parametric edits on the model after resolution (the modeling result of the edit in (a) is shown in (b), and similarly for (b), (c) and (d)).}\label{fig:result-sai-bracket-edit}
\end{figure}

Figs.~\ref{fig:result-sai-bracket}, \ref{fig:result-sai-bracket-resuelt} and \ref{fig:result-sai-bracket-edit} show an example integrating the four modules of GTI detection, GTI resolution, SAI detection, and SAI resolution. It is also based on a model downloaded from the GrabCAD library. Fig.~\ref{fig:result-sai-bracket}a shows the model shape updated after a direct edit that rotates its top faces. Fig.~\ref{fig:result-sai-bracket}b lists the updated model GCS, according to the new model shape. Fig.~\ref{fig:result-sai-bracket-resuelt} gives detected maximal well-constrained parts in the updated model GCS. There are in total 7 parts found. For each part, the belonging bounary faces are indicated by dark color, and the non-belonging ones are made transparent. Each part's face list have also been given at the bottom right of the figure. Figs.~\ref{fig:result-sai-bracket-edit}a - \ref{fig:result-sai-bracket-edit}d give modeling behavior under parametric edits after resolving all the GTIs and SAIs. As can be seen from the figure, the symmetric design intent between the two top features are successfully maintained, as expected. Changes made to one feature are reflected on the other.

\section{Conclusion and future work}
\label{sec:conclusion}
Modern CAD systems have two primary ways to do modeling: parametric and direct. Parametric/direct integration, which combines their complementary strengths, is emerging as an important development direction for CAD. This paper provided a detailed review on publicly reported methods in this direction from both academia and industry. The review showed that those methods are still at the early development stage, and that seamless parametric/direct integration still remains an open problem. This paper then presented an alternative approach, called variational direct modeling, to this problem. First, a problem analysis was conducted to identify the fundamental issues and challenges in parametric/direct integration. Subsequently, a framework was proposed to solve those challenges. Finally, effective algorithms to implement the framework's constituent modules were provided. 

In a nutshell, the underlying problem of parametric/direct integration is to handle the inconsistencies among geometric, topological, and constraint information in a model undergoing parametric/direct edits. Two critical information inconsistencies have been identified: GTIs and SAIs. The major issue of GTIs and SAIs is that there often exist many options for resolving them. The fundamental challenges lie in making systematic decisions among those options so that the following three requirements can be met: model validity (being solid and well-constrained), continuous shape variations, and minimal constraint changes. The proposed algorithms can work because they transformed these requirements into well-defined solid modeling operations (e.g., Booleans), known optimization problems (e.g., sparse recovery), and useful engineering notions (e.g, sensitivity).

The primary difference between the proposed method and existing methods is that it directly deals with the information inconsistencies in a model, while they try to indirectly handle those inconsistencies by converting direct edits into certain feature operations, which restricts either the parametric modeling capability or the direct modeling capability. Because of the direct strategy, the proposed method has the potential to achieve seamless integration of parametric and direct modeling. It should, however, be noted that this statement is not intended to imply that the method, in its current form, is already able to provide a complete solution. Seamless parametric/direct integration is a very challenging problem, requiring significant advances on robust solid modeling, fast geometric constraint solving, robust topological naming, and intelligent constraint recognition etc. This work only represents a further step towards seamless parametric/direct integration, and much more work remains to be done. Laying down the integration's fundamental challenges and providing a feasible framework for it are the major contributions of this work. With the challenges identified, the authors hope that they can attract more researchers to work on this very important research topic of parametric/direct integration. 

There are several important improvement directions for this work, including:
\begin{itemize}
    \itemsep0em    
    \item From a practical perspective, a more intuitive and interactive tool should be provided when presenting SAI resolution suggestions to the user. It is unreasonable to expect an average CAD user to understand the mathematical complexity and intricacies of over- and under-constrained parts, especially when the parts' sizes are large.
    \item The method used to prioritize options in SAI resolution may be augmented by artificial intelligence algorithms to provide more intelligent computer assistance and to decrease the need for manual intervention in SAI resolution. The method's current way of working could interrupt the design process when prompting the user to assist inconsistency resolution. This may affect the desinger's workflow. If augmented by intelligent decision-making where automatic resolution can work satisfactorily, at least requiring much less user assistance, the interruption issue can be mitigated significantly. It should, however, be noted that a 100\% automatic SAI resolution method seems to be impossible in the CAD domain since there are subjective aspects in engineering design.
    \item Another interesting augment that could be made is to improve the efficiency of GTI resolution. Boolean operations are currently used to carry out GTI resolution, which is compute-intensive. The computational load may become high when a complex model is considered. Parallel computing and localized Boolean operations (e.g., the method presented by Rossignac and Voelcker \cite{rossignac1988active}) may be helpful in this regard.
    \item Maintaining feature semantics in variational direct modeling is also a very important improvement direction. Particularly, the user can be intelligently assisted by information of whether or when an ongoing modeling operation will invalidate the model's feature semantics, e.g., a blind hole cannot be changed into a through hole. If the semantics are indeed broken by the user, a futher step in this direction is to automatically maintain the semantices of features, a problem similar to that considered in semantic feature modeling \cite{bidarra2000semantic}. The method presented there is thus very helpful.
    \item Additionally, including freeform surfaces into the current framework is of great interest. This not only introduces new challenging problems (e.g., preserving geometric continuity) and research opportunities but also increases this work's practical usability since many mechanical parts are composed of freeform surfaces, at least partially. Extending the present work to handling other operations, e.g., sweeping, is also of great interest. This can significantly improve the proposed framework's applicability and practical usefulness.
\end{itemize}

The parametric/direct integration method so developed advances the way solid models can be manipulated, in a more intuitive and intelligent way. Direct modeling is responsible for intuitive interaction between designers and solid models, while parametric modeling responsible for embedding design intent into solid models. These two benefits move solid modeling a further step towards the early, conceptual stages of design.

\section*{Acknowledgements}
This work has been funded by NSF of China (No. 62102355), NSF of Zhejiang Province (No. LQ22F020012), Key R\&D Program of Zhenjiang Province (No. 2022C01025), and a PhD fellowship from UBC.


\bibliographystyle{elsarticle-num}
\bibliography{Bibliography}

\begin{thebibliography}{10}
\expandafter\ifx\csname url\endcsname\relax
  \def\url#1{\texttt{#1}}\fi
\expandafter\ifx\csname urlprefix\endcsname\relax\def\urlprefix{URL }\fi
\expandafter\ifx\csname href\endcsname\relax
  \def\href#1#2{#2} \def\path#1{#1}\fi

\bibitem{li2020survey}
L.~Li, Y.~Zheng, M.~Yang, J.~Leng, Z.~Cheng, Y.~Xie, P.~Jiang, Y.~Ma, A survey
  of feature modeling methods: historical evolution and new development,
  Robotics and Computer-Integrated Manufacturing 61 (2020) 101851.

\bibitem{sapidis2007geometric}
N.~S. Sapidis, Geometric modeling of spatial constraints: objectives, methods
  and solid-modeling requirements, Computing 79~(2) (2007) 337--352.

\bibitem{hoffmann2005constraint}
C.~M. Hoffmann, Constraint-based computer-aided design, Tech. rep., Purdue
  University (2005).

\bibitem{requicha1985boolean}
A.~A. Requicha, H.~B. Voelcker, Boolean operations in solid modeling: Boundary
  evaluation and merging algorithms, Proceedings of the IEEE 73~(1) (1985)
  30--44.

\bibitem{requicha1982solid}
A.~A. Requicha, H.~B. Voelcker, Solid modeling: a historical summary and
  contemporary assessment, IEEE Computer Graphics and Applications 2~(02)
  (1982) 9--24.

\bibitem{requicha1983solid}
A.~A. Requicha, H.~B. Voelcker, Solid modeling: Current status and research
  directions, IEEE computer graphics and applications 3~(7) (1983) 25--37.

\bibitem{camba2016parametric}
J.~D. Camba, M.~Contero, P.~Company, Parametric cad modeling: an analysis of
  strategies for design reusability, Computer-Aided Design 74 (2016) 18--31.

\bibitem{gonzalez2017survey}
C.~Gonz{\'a}lez-Lluch, P.~Company, M.~Contero, J.~D. Camba, R.~Plumed, A survey
  on 3d cad model quality assurance and testing tools, Computer-Aided Design 83
  (2017) 64--79.

\bibitem{shah1998designing}
J.~J. Shah, Designing with parametric cad: classification and comparison of
  construction techniques, in: International Workshop on Geometric Modelling,
  Springer, 1998, pp. 53--68.

\bibitem{ElHani2012}
M.~A. {El Hani}, L.~Rivest, R.~Maranzana, {Product data reuse in product
  development: a practitioner's perspective}, in: IFIP International Conference
  on Product Lifecycle Management, Springer, 2012, pp. 243--256.

\bibitem{monedero2000parametric}
J.~Monedero, Parametric design: a review and some experiences, Automation in
  Construction 9~(4) (2000) 369--377.

\bibitem{Zou}
Q.~Zou, H.-Y. Feng, {Push-pull direct modeling of solid CAD models}, Advances
  in Engineering Software 127 (2019) 59--69.

\bibitem{qin2021automatic}
X.~Qin, Z.~Tang, S.~Gao, Automatic update of feature model after direct
  modeling operation, Computer-Aided Design and Applications 18 (2021) 170--85.

\bibitem{Fu2017}
J.~Fu, X.~Chen, S.~Gao, {Automatic synchronization of a feature model with
  direct editing based on cellular model}, Computer-Aided Design and
  Applications 14~(5) (2017) 680--692.

\bibitem{braid1975synthesis}
I.~C. Braid, The synthesis of solids bounded by many faces, Communications of
  the ACM 18~(4) (1975) 209--216.

\bibitem{voelcker1977geometric}
H.~B. Voelcker, A.~A. Requicha, Geometric modeling of mechanical parts and
  processes, Computer 10~(12) (1977) 48--57.

\bibitem{Shapiro2002}
V.~Shapiro, {Solid modeling}, in: Handbook of Computer Aided Geometric Design,
  North-Holland, 2002, pp. 473--518.

\bibitem{kyratzi2022integrated}
S.~Kyratzi, P.~Azariadis, Integrated design intent of 3d parametric models,
  Computer-Aided Design (2021) 103198.

\bibitem{Raghothama2002}
S.~Raghothama, V.~Shapiro, {Topological framework for part families}, in:
  Proceedings of the Seventh ACM Symposium on Solid Modeling and Applications,
  ACM, 2002, pp. 1--12.

\bibitem{bidarra2000semantic}
R.~Bidarra, W.~F. Bronsvoort, Semantic feature modelling, Computer-Aided Design
  32~(3) (2000) 201--225.

\bibitem{Tornincasa2010}
S.~Tornincasa, F.~D. Monaco, {The future and the evolution of CAD}, in:
  Proceedings of the 14th International Research/Expert Conference, 2010, pp.
  11--18.

\bibitem{Ault2016}
H.~Ault, A.~Phillips, {Direct modeling: easy changes in CAD}, in: Proceedings
  of the 70th ASEE EDGD Midyear Conference, 2016, pp. 99--106.

\bibitem{Grayer1980}
A.~R. Grayer, {Alternative approaches in geometric modelling}, Computer-Aided
  Design 12~(4) (1980) 189--192.

\bibitem{Rossignac1990}
J.~R. Rossignac, {Issues on feature-based editing and interrogation of solid
  models}, Computers and Graphics 14~(2) (1990) 149--172.

\bibitem{Stroud2006}
I.~Stroud, P.~C. Xirouchakis, {CAGD - Computer-aided gravestone design},
  Advances in Engineering Software 37~(5) (2006) 277--286.

\bibitem{bettig2011geometric}
B.~Bettig, C.~M~Hoffmann, Geometric constraint solving in parametric
  computer-aided design, Journal of Computing and Information Science in
  Engineering 11~(2) (2011).

\bibitem{nag2015methods}
A.~Nag, T.~D. Gallagher, J.~J. Dunne, Methods and systems for converting select
  features of a computer-aided design model to direct-edit features, uS Patent
  9,117,308 (2015).

\bibitem{Chad2008}
J.~Chad, H.~David, {Synchronous technology: the best of both worlds for
  engineering organizations}, Tech. rep., Aberdeen Group, Boston (2008).

\bibitem{lin1981variational}
V.~C. Lin, D.~C. Gossard, R.~A. Light, Variational geometry in computer-aided
  design, ACM SIGGRAPH 15~(3) (1981) 171--177.

\bibitem{chung2000framework}
J.~C. Chung, T.-S. Hwang, C.-T. Wu, Y.~Jiang, J.-Y. Wang, Y.~Bai, H.~Zou,
  Framework for integrated mechanical design automation, Computer-Aided Design
  32~(5-6) (2000) 355--365.

\bibitem{ushakov2008variational}
D.~Ushakov, Variational direct modeling: how to keep design intent in history
  free cad, Tech. rep., LEDAS Ltd (2008).

\bibitem{hoffman2001decomposition}
C.~M. Hoffman, A.~Lomonosov, M.~Sitharam, Decomposition plans for geometric
  constraint systems, part i: performance measures for cad, Journal of Symbolic
  Computation 31~(4) (2001) 367--408.

\bibitem{cordier2013inferring}
F.~Cordier, H.~Seo, M.~Melkemi, N.~S. Sapidis, Inferring mirror symmetric 3d
  shapes from sketches, Computer-Aided Design 45~(2) (2013) 301--311.

\bibitem{Mantyla1984a}
M.~Mantyla, {A note on the modeling space of Euler operators}, Computer Vision,
  Graphics, and Image Processing 26~(1) (1984) 45--60.

\bibitem{zou2020decision}
Q.~Zou, H.-Y. Feng, A decision-support method for information inconsistency
  resolution in direct modeling of cad models, Advanced Engineering Informatics
  44 (2020) 101087.

\bibitem{hu2017over}
H.~Hu, M.~Kleiner, J.-P. Pernot, Over-constraints detection and resolution in
  geometric equation systems, Computer-Aided Design 90 (2017) 84--94.

\bibitem{gonzalez2021constraint}
C.~Gonz{\'a}lez-Lluch, R.~Plumed, D.~P{\'e}rez-L{\'o}pez, P.~Company,
  M.~Contero, J.~D. Camba, A constraint redundancy elimination strategy to
  improve design reuse in parametric modeling, Computers in Industry 129 (2021)
  103460.

\bibitem{camba2015assessing}
J.~D. Camba, M.~Contero, Assessing the impact of geometric design intent
  annotations on parametric model alteration activities, Computers in Industry
  71 (2015) 35--45.

\bibitem{raghothama1998boundary}
S.~Raghothama, V.~Shapiro, Boundary representation deformation in parametric
  solid modeling, ACM Transactions on Graphics (TOG) 17~(4) (1998) 259--286.

\bibitem{lipp2014pushpull}
M.~Lipp, P.~Wonka, P.~M{\"u}ller, Pushpull++, ACM Transactions on Graphics
  33~(4) (2014) 1--9.

\bibitem{van2010tracking}
H.~A. Van~der Meiden, W.~F. Bronsvoort, Tracking topological changes in
  parametric models, Computer-Aided Geometric Design 27~(3) (2010) 281--293.

\bibitem{hidalgo2012computing}
M.~Hidalgo, R.~Joan-Arinyo, Computing parameter ranges in constructive
  geometric constraint solving: implementation and correctness proof,
  Computer-Aided Design 44~(7) (2012) 709--720.

\bibitem{zou2021robust}
Q.~Zou, H.-Y. Feng, A robust direct modeling method for quadric b-rep models
  based on geometry--topology inconsistency tracking, Engineering with
  Computers (2021) 1--16.

\bibitem{hu2021geometric}
H.~Hu, M.~Kleiner, J.-P. Pernot, C.~Zhang, Y.~Huang, Q.~Zhao, S.~Yeung,
  Geometric over-constraints detection: a survey, Archives of Computational
  Methods in Engineering 28~(7) (2021) 4331--4355.

\bibitem{dufourd1998geometric}
J.-F. Dufourd, P.~Mathis, P.~Schreck, Geometric construction by assembling
  solved subfigures, Artificial Intelligence 99~(1) (1998) 73--119.

\bibitem{owen1991algebraic}
J.~C. Owen, Algebraic solution for geometry from dimensional constraints, in:
  Proceedings of the first ACM symposium on Solid modeling foundations and
  CAD/CAM applications, 1991, pp. 397--407.

\bibitem{bouma1995geometric}
W.~Bouma, I.~Fudos, C.~Hoffmann, J.~Cai, R.~Paige, Geometric constraint solver,
  Computer-aided design 27~(6) (1995) 487--501.

\bibitem{fudos1997graph}
I.~Fudos, C.~M. Hoffmann, A graph-constructive approach to solving systems of
  geometric constraints, ACM Transactions on Graphics (TOG) 16~(2) (1997)
  179--216.

\bibitem{gao2002solving}
X.-S. Gao, C.~M. Hoffmann, W.-Q. Yang, Solving spatial basic geometric
  constraint configurations with locus intersection, in: Proceedings of the
  seventh acm symposium on solid modeling and applications, 2002, pp. 95--104.

\bibitem{barford1987graphical}
L.~A. Barford, A graphical, language-based editor for generic solid models
  represented by constraints, Ph.D. thesis, Cornell University (1987).

\bibitem{serrano1987constraint}
D.~Serrano, Constraint management in conceptual design, Ph.D. thesis,
  Massachusetts Institute of Technology (1987).

\bibitem{ait2014reduction}
S.~Ait-Aoudia, R.~Jegou, D.~Michelucci, Reduction of constraint systems, in:
  Proceedings of Compugraphics, 1993, pp. 83–--92.

\bibitem{latham1996connectivity}
R.~S. Latham, A.~E. Middleditch, Connectivity analysis: a tool for processing
  geometric constraints, Computer-Aided Design 28~(11) (1996) 917--928.

\bibitem{hoffmann1997finding}
C.~M. Hoffmann, A.~Lomonosov, M.~Sitharam, Finding solvable subsets of
  constraint graphs, in: International Conference on Principles and Practice of
  Constraint Programming, Springer, 1997, pp. 463--477.

\bibitem{hoffmann2004making}
C.~M. Hoffmann, M.~Sitharam, B.~Yuan, Making constraint solvers more usable:
  overconstraint problem, Computer-Aided Design 36~(4) (2004) 377--399.

\bibitem{michelucci2006geometric}
D.~Michelucci, S.~Foufou, Geometric constraint solving: the witness
  configuration method, Computer-Aided Design 38~(4) (2006) 284--299.

\bibitem{thierry2011extensions}
S.~E. Thierry, P.~Schreck, D.~Michelucci, C.~F{\"u}nfzig, J.-D. G{\'e}nevaux,
  Extensions of the witness method to characterize under-, over-and
  well-constrained geometric constraint systems, Computer-Aided Design 43~(10)
  (2011) 1234--1249.

\bibitem{zou2019variational}
Q.~Zou, H.-Y. Feng, Variational b-rep model analysis for direct modeling using
  geometric perturbation, Journal of Computational Design and Engineering 6~(4)
  (2019) 606--616.

\bibitem{fornasier2015compressive}
M.~Fornasier, H.~Rauhut, Compressive sensing, Handbook of Mathematical Methods
  in Imaging (2015) 187--229.

\bibitem{Osher2014}
S.~Osher, W.~Yin, {Sparse recovery via l1 and L1 optimization}, Tech. rep.,
  University of California, Los Angeles (2014).

\bibitem{murugappan2009towards}
S.~Murugappan, S.~Sellamani, K.~Ramani, Towards beautification of freehand
  sketches using suggestions, in: Proceedings of the 6th Eurographics Symposium
  on Sketch-Based Interfaces and Modeling, 2009, pp. 69--76.

\bibitem{mills2001estimate}
B.~Mills, F.~Langbein, A.~Marshall, R.~Martin, Estimate of frequencies of
  geometric regularities for use in reverse engineering of simple mechanical
  components, Submitted to International Journal of Shape Modeling (2001).

\bibitem{martinez2005constraint}
M.~L. Mart{\'\i}nez, J.~F{\'e}lez, A constraint solver to define correctly
  dimensioned and overdimensioned parts, Computer-Aided Design 37~(13) (2005)
  1353--1369.

\bibitem{zou2007constraint}
H.~Zou, Y.~Lee, Constraint-based beautification and dimensioning of 3d
  polyhedral models reconstructed from 2d sketches, Computer-Aided Design
  39~(11) (2007) 1025--1036.

\bibitem{langbein2004choosing}
F.~C. Langbein, A.~D. Marshall, R.~R. Martin, Choosing consistent constraints
  for beautification of reverse engineered geometric models, Computer-Aided
  Design 36~(3) (2004) 261--278.

\bibitem{li2011globfit}
Y.~Li, X.~Wu, Y.~Chrysathou, A.~Sharf, D.~Cohen-Or, N.~J. Mitra, Globfit:
  Consistently fitting primitives by discovering global relations, ACM
  Transactions on Graphics 30~(4) (2011) 52:1--52:12.

\bibitem{hillyard1978analysis}
R.~Hillyard, I.~Braid, Analysis of dimensions and tolerances in computer-aided
  mechanical design, Computer-Aided Design 10~(3) (1978) 161--166.

\bibitem{rossignac1988active}
J.~R. Rossignac, H.~B. Voelcker, Active zones in csg for accelerating boundary
  evaluation, redundancy elimination, interference detection, and shading
  algorithms, ACM Transactions on Graphics (TOG) 8~(1) (1988) 51--87.

\end{thebibliography}

\end{document}